\documentclass[prl,showpacs,superscriptaddress,twocolumn]{revtex4-1}
\usepackage{amsfonts}
\usepackage{amssymb}
\usepackage{amsmath}
\usepackage{graphicx}
\usepackage{epsfig}
\usepackage{color}

\begin{document}

\title{Incident Direction Independent Wave Propagation and Unidirectional
Lasing}
\author{L. Jin}
\email{jinliang@nankai.edu.cn}
\author{Z. Song}
\affiliation{School of Physics, Nankai University, Tianjin 300071, China}

\begin{abstract}
We propose an incident direction independent wave propagation generated by
properly assembling different unidirectional destructive interferences
(UDIs), which is a consequence of the appropriate match between synthetic
magnetic fluxes and the incident wave vector. Single-direction lasing at
spectral singularity is feasible without introducing nonlinearity. UDI
allows unidirectional lasing and unidirectional perfect absorption; when
they are combined in a parity-time symmetric manner, the spectral
singularities vanish with bounded reflections and transmissions.
Furthermore, the simultaneous unidirectional lasing and perfect absorption
for incidences from opposite directions is created. Our findings provide
insights into light control and may shed light on the explorations of
desirable functionality in fundamental research and practical applications.
\end{abstract}

\maketitle

\emph{Introduction}.---Parity-time ($\mathcal{PT}$) symmetry has been
theoretically and experimentally investigated in a variety of non-Hermitian
systems~\cite%
{Bender98,Ali02,Jones,Ruschhaupt,El,Musslimani,Klaiman,LJIN09,ZLin,SRotter,VVK,HJing,LGePRX,ZhuPRX,Chang,Alu,AAS,AGuo,CE,BP,ZZhang,PTRev}%
, non-Hermiticity controls the exact and broken $\mathcal{PT}$-symmetric
phases~\cite{AGuo,CE,BP,ZZhang,PTRev}. The phase transition points are
exceptional points~\cite{Dembowski,Wunner,Rotter,Uzdin,Heiss12} utilized for
sensing enhancement~\cite{Wiersig,Wiersig2016,ZPLiu,EP2Sensing,EP3Sensing}.
The topologies of exceptional points are distinct~\cite%
{Zhen,Menke,Doppler,Xu,KDingPRX}. Spectral singularities (SS) in scattering
systems belong to another type of non-Hermitian singularities, at which
eigenstate completeness is spoiled~\cite{Ali,Ali13}; incident waves from
opposite directions at an appropriate phase match are perfectly absorbed in
a coherent perfect absorber~\cite%
{CPA,LonghiCPA,Chong11,Wan,Sun,Wong,HangCPA,LGePRA17,CPAREV}.

Non-Hermitian character causes unidirectionality~\cite%
{PengPNAS,Peschel,LFeng,USS,LJINSR,UPA}, the fundamental mechanism of which
differs from that created by chiral light-matter interaction~\cite%
{TR,TRPRA,PL,PRA93,PZ}. Typical phenomena include unidirectional
reflectionlessness~\cite{Peschel,LFeng} and unidirectional spectral
singularity that allows unidirectional perfect absorption (UPA) and
unidirectional lasing (UL)~\cite{USS,UPA}; however, the transmissions there
protected by symmetry are reciprocal~\cite{XYin,Jalas,MugaPR,PTScattAnnPhys}%
. Nonreciprocal transmission is indispensable for optical information
processing. Nonreciprocity, implemented via magneto-optic effect~\cite{LBi}
and optical nonlinearity~\cite{LFan}, has been created based on various
strategies in linear and magnetic-free devices~\cite{ZYu,DWWang}, in
single-photon level~\cite{BDayan2014,BDayan2016}, and even in acoustics~\cite%
{Fleury2014}. Benefitted from synthetic magnetic flux realized for photons~%
\cite{Fang12,Tzuang14,LLu,ELi,SLonghiOL14,Hafezi2011,Hafezi,HafeziIJMPB,NP13}
and progress in non-Hermitian physics~\cite{PTRev}, non-Hermiticity
associated with synthetic magnetic flux induces nonreciprocal transmission
in linear photonic lattices~\cite{LonghiOL,LXQ,JLNJP}.

In this Letter, we propose an incident direction independent wave
propagation, which is an extraordinary asymmetry in both \emph{reflectivity}
and \emph{transmittivity} that stemmed from the unidirectional destructive
interference (UDI); single-direction lasing occurs at the SS, where an
incidence from either side induces a UL toward the same direction. The
appropriate match between synthetic magnetic fluxes in different functional
UDIs creates many intriguing unidirectional phenomena beyond one-way
propagation, UPA, and UL~\cite{JLNJP}. With judiciously chosen synthetic
magnetic fluxes, detunings, and the gain or loss of side-coupled resonators,
the SS exhibits UPA or UL; their coincidence leads to an SS elimination
under $\mathcal{PT}$ symmetry. Furthermore, a simultaneous unidirectional
lasing and perfect absorption for incidences from opposite sides is feasible
by assembling different unidirectional elements. These novel wave
propagation phenomena facilitate various applications of non-Hermitian
metamaterials without introducing nonlinearity~\cite%
{SLonghi15,Lepri,NL,VVKRMP}.

\begin{figure}[tb]
\includegraphics[ bb=0 0 540 190, width=8.8 cm, clip]{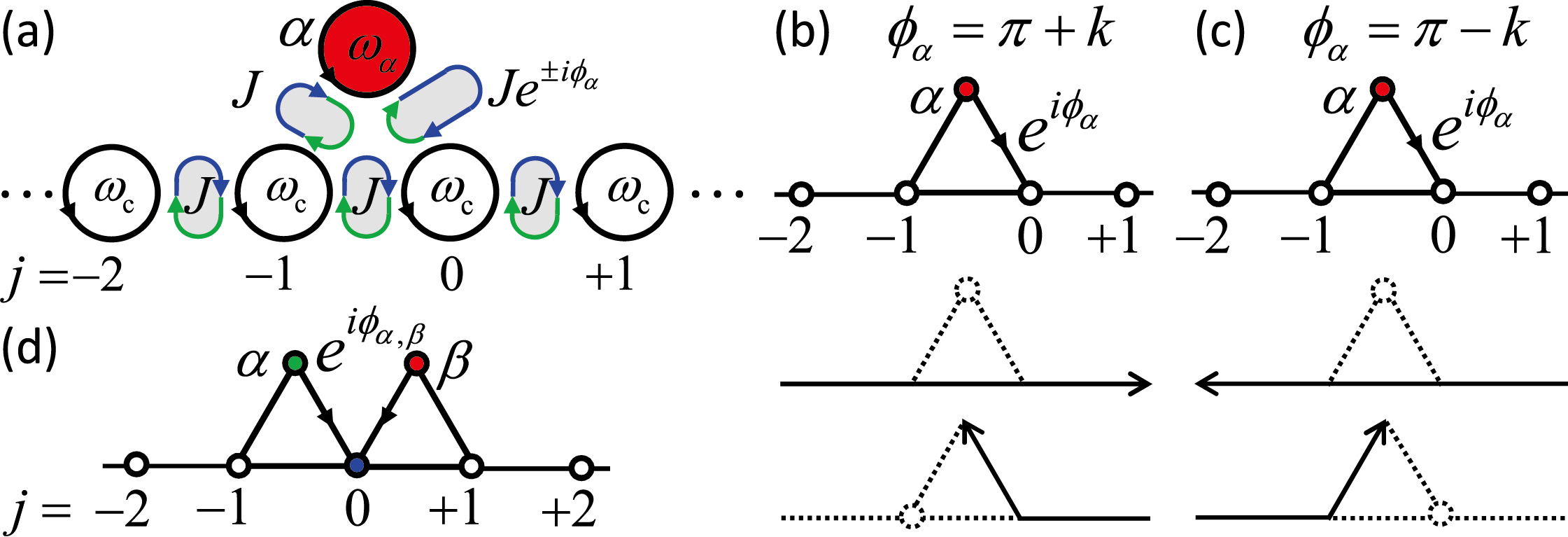}
\caption{(a) Schematic of uniform resonator chain with one side-coupled resonator, $\omega_{\alpha}=\omega_{\rm{c}}+V_{\alpha}$.
The blue and green arrows indicate the optical path lengths for counterclockwise mode (black arrow) photons tunneling between resonators in opposite directions. (b) Right-UPA: right perfect absorption and left resonant transmission. (c) Left-UPA: left perfect absorption and right resonant transmission. The upper
panels illustrate UPA configurations with opposite synthetic magnetic fluxes, $k$ is the incident wave vector and $V_{\alpha}=-e^{ik}$; the central panels illustrate destructive interference at the side-coupled resonator $\alpha$ from one incident direction; and the lower panels illustrate perfect absorption from the opposite incident direction, the dotted line indicates the equivalently decoupled part. (d) Schematic of uniform resonator chain with two side-coupled resonators. $j$ is the resonator index, $\omega_{\alpha, 0, \beta}=\omega_{\rm{c}}+V_{\alpha, 0, \beta}$.}
\label{fig1}
\end{figure}

\emph{Model}.---We consider a one-dimensional uniformly coupled passive
resonator chain in Fig.~\ref{fig1}(a) with identical resonant frequency $%
\omega _{\mathrm{c}}$; the primary resonators (round-shaped) are
evanescently coupled through the auxiliary resonators (stadium-shaped),
which are antiresonant with the primary resonators. The coupling strength $%
J=1$ is set to be unity. A resonator is side-coupled to the resonator chain
with one asymmetric coupling, which is introduced through the optical path
length difference $2\Delta x$ in the tunneling between resonators $\alpha $
and $0$~\cite{Hafezi}. An extra direction-dependent phase factor $e^{\pm
i\phi _{\alpha }}$ is equivalently induced in the effective coupling between
resonators $\alpha$ and $0$, resulting in a synthetic magnetic flux $\phi
_{\alpha }=2\pi \Delta x/\lambda \equiv (-e/\hbar )\oint \vec{A}_{\alpha
}\cdot d\vec{l}$ in the triangle that associated with a vector potential $%
\vec{A}_{\alpha }$~\cite{LLu,ELi,NP13}, where $\lambda $ is the resonant
wavelength~\cite{HafeziIJMPB}. The resonators support the counterclockwise
(CCW) and clockwise (CW) modes, they experience opposite synthetic magnetic
fluxes. The CCW mode is analyzed without loss of generality, the CW mode is
discussed as well.

The frequency detuning (net gain or loss) of resonator $\alpha $ is
represented by the real (imaginary) part of $V_{\alpha }$. The loss is the
dissipation $\omega _{\mathrm{c}}/(2Q_{\alpha})$ caused by the interaction
between resonator $\alpha $ and its environment, where $Q_{\alpha}$ is the
quality factor. The gain is induced by pumping the ions doped in the
resonator~\cite{Chang,BP}. Assuming a weak pump that far away from gain
saturation with white noise~\cite{Haken,SYZhu}, both processes are Markovian
and characterized by Lindblad master equation~\cite{Breuer,Scully}, where
the gain and loss are modeled by constants~\cite%
{BP,USS,FengScience,JLNC,Chang}.

We consider an ideal case of resonators with negligible backscattering~\cite%
{Supplementary}. In the coupled-mode theory~\cite{Haus,Joannopoulos}, the
equation of motion for resonators $j\neq -1,0$, and $\alpha $ in the system
illustrated in Fig.~\ref{fig1}(a) is
\begin{equation}
i\frac{\mathrm{d}\psi _{j}}{\mathrm{d}t}=\omega _{\mathrm{c}}\psi _{j}-\psi
_{j-1}-\psi _{j+1},  \label{1}
\end{equation}%
otherwise,%
\begin{eqnarray}
&&i\frac{\mathrm{d}\psi _{-1}}{\mathrm{d}t}=\omega _{\mathrm{c}}\psi
_{-1}-\psi _{-2}-\psi _{0}-\psi _{\alpha },  \label{2} \\
&&i\frac{\mathrm{d}\psi _{\alpha }}{\mathrm{d}t}=\left( \omega _{\mathrm{c}%
}+V_{\alpha }\right) \psi _{\alpha }-\psi _{-1}-e^{-i\phi _{\alpha }}\psi
_{0},  \label{3} \\
&&i\frac{\mathrm{d}\psi _{0}}{\mathrm{d}t}=\omega _{\mathrm{c}}\psi
_{0}-\psi _{-1}-\psi _{1}-e^{i\phi _{\alpha }}\psi _{\alpha },  \label{4}
\end{eqnarray}%
where $\psi _{j}=f_{j}e^{-i\omega t}$ is the field amplitude and $f_{j}$ is
the steady-state wave function of resonator $j$ in the elastic scattering
process with $f_{j}=Ae^{ikj}+Be^{-ikj}$ ($j<0$) and $f_{j}=Ce^{ikj}+De^{-ikj}
$ ($j>0$). $k\in \lbrack -\pi ,\pi ]$ is the dimensionless Bloch wave vector~%
\cite{MugaPR,PTScattAnnPhys,USS,LonghiOL}. The resonator chain supported
dispersion relation is $\omega =\omega _{\mathrm{c}}-2J\cos k$~\cite%
{USS,Hafezi}, a resonant incidence with frequency $\omega _{\mathrm{c}}$ has
the wave vector $k=\pi /2$.

\emph{Unidirectional destructive interference}.---Synthetic magnetic flux
affects the interference, breaks the system symmetry and the reciprocity of
transmission~\cite{Supplementary}. When synthetic magnetic flux matches the
incident wave vector $\phi _{\alpha }=\pi \pm k$, incidences from both sides
are reflectionless. The side-coupled resonator $\alpha $ is equivalently
isolated for the left (right) incidence because of the destructive
interference induced by $\phi _{\alpha }=\pi +k$ ($\phi _{\alpha }=\pi -k$);
the wave function at resonator $\alpha $ is zero. For the right (left)
incidence, the wave function of resonator $\alpha $ does not vanish, it
varies according to $V_{\alpha }$ and affects the right (left) transmission.

The reflection and transmission coefficients are $r_{\mathrm{L}}=B/A$, $t_{%
\mathrm{L}}=C/A$ for the left incidence ($D=0$); and $r_{\mathrm{R}}=C/D$, $%
t_{\mathrm{R}}=B/D$ for the right incidence ($A=0$). The scattering matrix
characterizes the relationship between input and output~\cite{Jalas,CPAREV}
\begin{equation}
\left(
\begin{array}{c}
B \\
C%
\end{array}%
\right) =S\left(
\begin{array}{c}
A \\
D%
\end{array}%
\right) .  \label{S}
\end{equation}%
Here, the scattering matrix is asymmetric~\cite{Supplementary}
\begin{equation}
S=\left(
\begin{array}{cc}
r_{\mathrm{L}} & t_{\mathrm{R}} \\
t_{\mathrm{L}} & r_{\mathrm{R}}%
\end{array}%
\right) =\left(
\begin{array}{cc}
0 & \left( e^{ik}+V_{\alpha }\right) /\left( e^{-ik}+V_{\alpha }\right) \\
1 & 0%
\end{array}%
\right) ,  \label{S2}
\end{equation}%
induced by the asymmetric coupling. In Hermitian systems (real $V_{\alpha }$%
),
the transmittivity is $|t_{\mathrm{L}}|^2= |t_{\mathrm{R}}|^2=1$; $f_{\alpha}
$ vanishes for the left incidence, but is nonzero for the right incidence.
In non-Hermitian systems (complex $V_{\alpha }$), the transmittivity is
asymmetric, $|t_{\mathrm{L}}|^2\neq |t_{\mathrm{R}}|^2$.

At $\phi _{\alpha }=\pi +k$, the scattering of left incidence with wave
vector $k$ is fixed even though $V_{\alpha }$ varies; the perfect absorption
occurs for the right incidence when $V_{\alpha }=-e^{ik}$ [Fig.~\ref{fig1}%
(b)]; in contrast to an isolator~\cite{Jalas}, the scattering matrix $%
S=\left(
\begin{array}{cc}
0 & 0 \\
1 & 0%
\end{array}%
\right) $ is for the CCW mode and its transpose $S^{\mathrm{T}}$ is for the
CW mode~\cite{Supplementary}. The wave function for the right incidence
consists of an incoming wave that is completely absorbed at resonator $%
\alpha $ without reflection, i.e., $f_{j}=0$ ($j<0$), $f_{j}=e^{-ikj}$ ($%
j\geqslant 0$), and $f_{\alpha }=-1$; $r_{\mathrm{L}}=r_{\mathrm{R}}=t_{%
\mathrm{R}}=0$ and $\left\vert t_{\mathrm{L}}\right\vert =1$, behaving
similarly in a coupled helical waveguide design~\cite{LonghiOL} but
differently in the transmissionless-UPA~\cite{UPA}. When the synthetic
magnetic flux is opposite$\ \phi _{\alpha }=\pi -k$, the wave function is a
left-right mirror reflection of that at $\phi _{\alpha }=\pi +k$, where
resonator $\alpha $ is isolated for the right incidence [Fig.~\ref{fig1}%
(c)]. A right-UPA for the CCW mode is a left-UPA for the CW mode, UPA
realizes chiral mode isolation~\cite{ZYu,LBi,LFan}.

UL occurs at $V_{\alpha }=-e^{-ik}$ when $t_{\mathrm{R}}$ diverges~\cite%
{Ali,USS,Ramezani2014}, where resonator $\alpha $ has an equal amount of
gain in contrast to the UPA. The wave function for the right incidence
consists of an outgoing wave that satisfies the boundary condition of lasing
without any injection, i.e., $f_{j}=e^{-ikj}$ ($j<0$), $f_{j}=0$ ($%
j\geqslant 0$), and $f_{\alpha }=1$.

\emph{Incident direction independent wave propagation}.---UDI facilitates
the design of optical control devices, leading to an isolation of
side-coupled resonator in one incident direction and providing an
opportunity to unidirectionally manipulate the waves. When assembling one
more side-coupled resonator, the cooperation between synthetic magnetic
fluxes in UDIs enriches the intriguing asymmetric dynamics~\cite{JLNJP}.

The equations of motion for the configuration of two side-coupled resonators
[Fig.~\ref{fig1}(d)] are Eqs.~(\ref{1}-\ref{3}) and
\begin{eqnarray}
&&i\frac{\mathrm{d}\psi _{0}}{\mathrm{d}t}=\left( \omega _{\mathrm{c}%
}+V_{0}\right) \psi _{0}-\psi _{-1}-\psi _{1}-e^{i\phi _{\alpha }}\psi
_{\alpha }-e^{i\phi _{\beta }}\psi _{\beta }, \\
&&i\frac{\mathrm{d}\psi _{\beta }}{\mathrm{d}t}=\left( \omega _{\mathrm{c}%
}+V_{\beta }\right) \psi _{\beta }-\psi _{1}-e^{-i\phi _{\beta }}\psi _{0},
\\
&&i\frac{\mathrm{d}\psi _{1}}{\mathrm{d}t}=\omega _{\mathrm{c}}\psi
_{1}-\psi _{0}-\psi _{2}-\psi _{\beta },
\end{eqnarray}
where the real (imaginary) parts of $V_{\beta }$ and $V_{0}$ represent the
frequency detunings (net gains or losses) of resonators $\beta $ and $0$,
respectively.

\begin{figure}[t]
\centering
\includegraphics[ bb=10 0 240 100, width=8.8 cm, clip]{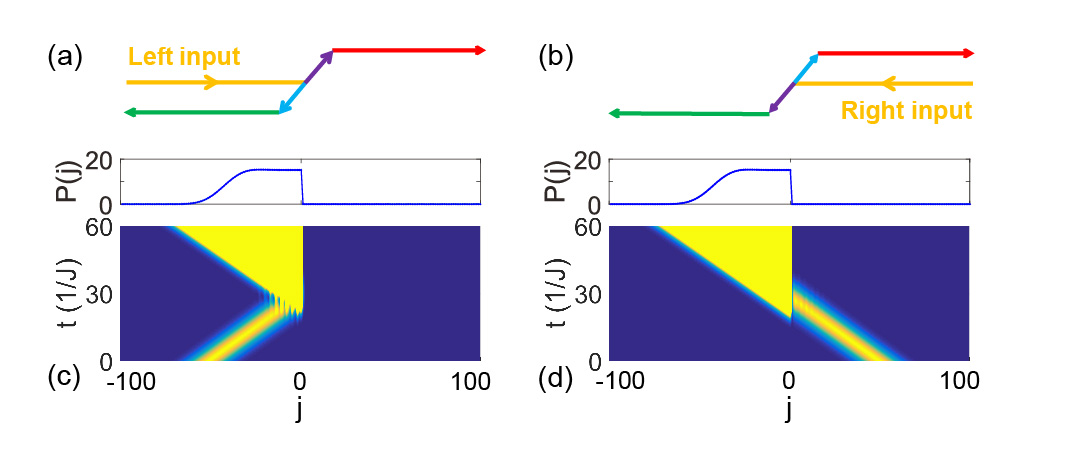}
\caption{(a, b) Schematic of the incident direction independent wave propagation. (c, d)
Simulations of single-direction lasing at $V_{\protect\alpha }=-e^{-ik}$, $V_{\protect\beta }=-e^{ik}$ in Fig.~\ref{fig1}(d). The Gaussian wave packet is $\left\vert \Psi \left( 0,j\right) \right\rangle =(\protect\sqrt{\protect\pi }/\protect\sigma )^{-1/2}\sum_{j}e^{-(\protect\sigma ^{2}/2)\left( j-N_{\mathrm{c}}\right) ^{2}}e^{ik_{\mathrm{c}}j}\left\vert j\right\rangle $,
centered at $N_{\mathrm{c}}$, where $k_{\mathrm{c}}=\protect\pi /3$ is the
wave vector, $\protect\sigma =0.1$, and $j$
is the resonator index. The resonator chain is cut at $j=\pm 100$. The blue curves in (c, d) depict
the wave intensities $P\left( j\right) =|\Psi \left( t,j\right) |^{2}$ at
time $t=60/J$. Other parameters are $\protect\phi _{\protect\alpha }=\protect\phi _{\protect\beta }=\protect\pi +k$, $V_{0}=-2i\sin k$, and $k=\protect\pi /3$.}
\label{fig2}
\end{figure}

At fixed synthetic magnetic fluxes $\phi _{\alpha }=\phi _{\beta }=\pi +k$,
Figs.~\ref{fig2}(a) and~\ref{fig2}(b) illustrate the wave propagation for
opposite incidences. In Fig.~\ref{fig2}(a), the wave resonantly transmits at
resonator $\alpha $ because of the destructive interference. Then, the
transmitted wave is scattered at resonator $0$, corresponding reflection and
transmission coefficients are $r_{0}=V_{0}/(2i\sin k-V_{0})$ and $%
t_{0}=r_{0}+1$~\cite{Kim}. The reflected portion (cyan arrow) passes through
resonator $\alpha $ from the right side without reflection (green arrow) and
forms the left reflection; the transmitted portion (purple arrow) passes
through resonator $\beta $ from the left side without reflection (red arrow)
and forms the left transmission. The green (red) arrow represents the
reflectionless transmission modulated by resonator $\alpha $ ($\beta $). The
reflection and transmission coefficients are~\cite{Supplementary}:%
\begin{eqnarray}
r_{\text{\textrm{L}}} &=&r_{0}\frac{e^{ik}+V_{\alpha }}{e^{-ik}+V_{\alpha }}%
,t_{\mathrm{L}}=t_{0}\frac{e^{ik}+V_{\beta }}{e^{-ik}+V_{\beta }};
\label{DSL} \\
r_{\mathrm{R}} &=&r_{0}\frac{e^{ik}+V_{\beta }}{e^{-ik}+V_{\beta }},t_{%
\mathrm{R}}=t_{0}\frac{e^{ik}+V_{\alpha }}{e^{-ik}+V_{\alpha }}.  \label{DSR}
\end{eqnarray}


The synthetic magnetic fluxes allow that the propagating waves in the left
(right) chain after scattering are $V_{\beta (\alpha )}$ independent,
regardless of the incident direction. This enables an incident direction
independent wave propagation when $\left\vert r_{0}\right\vert =\left\vert
t_{0}\right\vert $ (requiring $\left\vert 2\sin k/V_{0}\right\vert =1$): the
left reflectivity and transmittivity equal to the right transmittivity and
reflectivity, respectively.
\begin{equation}
\left\vert r_{\mathrm{L}}\right\vert ^2=\left\vert t_{\mathrm{R}}\right\vert
^2 ,\left\vert t_{\mathrm{L}}\right\vert ^2=\left\vert r_{\mathrm{R}%
}\right\vert ^2.
\end{equation}%
The left- and right-going propagating wave \emph{intensities} after
scattering are identical for incidences impinging from both directions and
are separately tuned by $V_{\alpha }$ and $V_{\beta }$, respectively. The
wave impinging from either direction is equally divided at resonator $0$
with $\left\vert r_{0}\right\vert =\left\vert t_{0}\right\vert =\sqrt{%
1/\left( 2-2\sin \varphi \right) }$ when $V_{0}=2e^{i\varphi }\sin k$.

The incident direction independent wave propagation occurs at
\begin{equation}
\phi _{\alpha }=\phi _{\beta }=\pi +k,V_{0}=2e^{i\varphi }\sin k.
\end{equation}

Resonator $\beta $ ($\alpha $) being $V_{\beta (\alpha )}=-e^{ik}$ induces a
perfect absorption of the right- (left-) going propagating wave, except that
the system is at the SS when $V_{0}=2i\sin k$, where lasing is
bidirectional; or when $V_{\alpha (\beta )}=-e^{-ik}$, where the lasing is
unidirectional toward the left (right) and the propagating wave in the right
(left) chain vanishes after scattering, e.g., UL occurs when $V_{\alpha
}=-e^{-ik}$, $V_{\beta }=-e^{ik}$, and $V_{0}=-2i\sin k$; the lasing wave is
emitted from the left chain, independent of the incident direction; and the
right-going propagating wave vanishes, being absorbed at resonator $\beta $.
The reflection and transmission coefficients satisfy
\begin{equation}
\left\vert r_{\mathrm{L}}\right\vert =\left\vert t_{\mathrm{R}}\right\vert
\rightarrow \infty ,t_{\mathrm{L}}=r_{\mathrm{R}}=0.
\end{equation}%
The wave function is $f_{j}=e^{-ikj}$ ($j<0$), $f_{j}=0$ ($j\geqslant 0$), $%
f_{\alpha }=1$, and $f_{\beta }=0$; consisting of the outgoing wave in the
left chain. For an incidence from either direction, the lasing wave is
emitted in a single direction: leftward. The simulation of single-direction
lasing is depicted in Figs.~\ref{fig2}(c) and~\ref{fig2}(d). A Gaussian wave
packet excites the wave emission, characterized by a Gaussian error function
and its intensity increases linearly~\cite{WP}. At $V_{0}=2i\sin k$, lasing
is bidirectional when $V_{\alpha ,\beta }\neq -e^{ik} $; the wave emission
is symmetric if $V_{\alpha ,\beta }$ is real, but asymmetric if complex $%
V_{\alpha }\neq V_{\beta }$; when $V_{\alpha (\beta )}=-e^{ik}$, the wave
emission is absorbed at resonator $\alpha $ ($\beta $) and the lasing
becomes unidirectional with vanishing emission toward the left (right). When
$V_{\alpha }=V_{\beta }=-e^{-ik}$, lasing is bidirectional and $V_{0}$
controls the asymmetry of lasing amplitude.

Another intriguing application is the one-way propagation at $V_{\alpha
}=-\cos k+3^{\pm 1}i\sin k$, $V_{\beta }=-e^{ik}$, and $V_{0}=-2i\sin k$
previously designed using different strategies~\cite{JLNJP,SLonghi15,Lepri},
rectifying waves with $\left\vert r_{\mathrm{L}}\right\vert =\left\vert t_{%
\mathrm{R}}\right\vert =1$ and $t_{\mathrm{L}}=r_{\mathrm{R}}=0$.

\begin{figure}[t]
\centering\includegraphics[ bb=0 0 520 130, width=8.8 cm, clip]{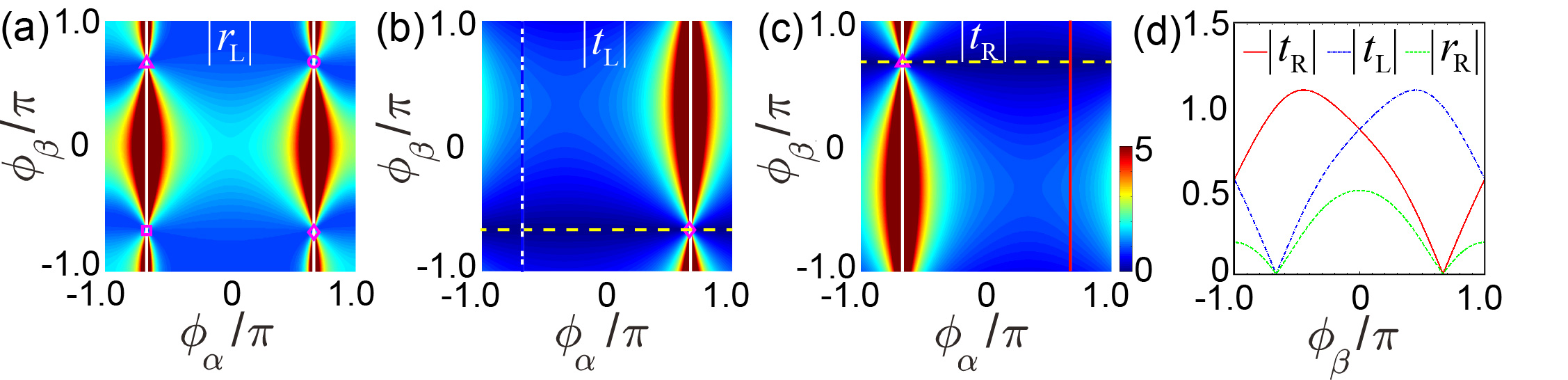}
\caption{Density plots of (a) $|r_{\mathrm{L}}|$, (b) $|t_{\mathrm{L}}|$, and
(c) $|t_{\mathrm{R}}|$ as functions of $\protect\phi _{\protect\alpha }$ and $\protect\phi _{\protect\beta }$. $r_{\mathrm{R}}=0$ for $\left\vert \protect\phi _{\protect\alpha }\right\vert \neq 2\protect\pi /3$; (d) $|t_{\mathrm{L}}|
$ ($|t_{\mathrm{R}}|$) for $\protect\phi _{\protect\alpha }=-2\protect\pi /3$ $(2\protect\pi /3)$ and $|r_{\mathrm{R}}|$ at when $r_{\mathrm{L}}$ diverges. Solid white (dashed yellow) lines indicate the divergence (zero) at the SS. SS coincide with unity values at the marked points in (b, c). Color bars are all in (c), $|r_{\mathrm{L}}|$ and $|t_{\mathrm{L,R}}|$ are cut to 5. The parameters are $V_{\protect\alpha }=-e^{-i\pi/3}$, $V_{\protect\beta }=-e^{i\pi/3}$,
$V_{0}=0$, and the
incident wave vector is $k=\protect\pi /3$.} \label{figPT}
\end{figure}

For $V_{0}=0$, the scattering is both sides reflectionless when $V_{\alpha
,\beta }\neq -e^{-ik}$. The left (right) transmission depends on $V_{\beta
(\alpha )}$. When $V_{\alpha }=V_{\beta }=-e^{-ik}$, the transmission
coefficients diverge with finite reflections. When $V_{\alpha }=V_{\beta
}=-e^{ik}$, the system completely absorbs the incidence from either
direction without reflection. When $V_{\alpha }=-e^{-ik}$ and $V_{\beta
}=-e^{ik}$, UL occurs with $\left\vert r_{\mathrm{L}}\right\vert =1$, $\ t_{%
\mathrm{L}}=r_{\mathrm{R}}=0$, and $\left\vert t_{\mathrm{R}}\right\vert
\rightarrow \infty $ (square in Fig.~\ref{figPT}).

$\mathcal{PT}$\emph{-symmetric side-coupled resonators}.---UPA and UL form a
Hermitian conjugation pair. Their SS coincide and vanish at series
connection of the two structures in a $\mathcal{PT}$-symmetric manner. The
wave emission is absorbed and leaves finite scattering intensities.

Figure~\ref{figPT} depicts $|r_{\mathrm{L,R}}|$ and $|t_{\mathrm{L,R}}|$ for
$k=\pi/3$~\cite{Supplementary}. $|\phi _{\alpha (\beta )}|=2\pi /3$ produces
a wave emission (absorption). $\left\vert r_{\mathrm{L}}\right\vert =1$ at $%
\left\vert \phi _{\beta }\right\vert =2\pi /3$ and $\left\vert r_{\mathrm{L}%
}\right\vert \rightarrow \infty $ at $|\phi _{\alpha }|=2\pi /3$ and $|\phi
_{\beta }|\neq 2\pi /3$ [Fig.~\ref{figPT}(a)]. $V_{\beta }=-e^{ik} $ results
in $f_{0}=0$ for the right incidence except when $r_{\mathrm{L}}$ diverges;
consequently, $r_{\mathrm{R}}=0$. Figures~\ref{figPT}(b) and~\ref{figPT}(c)
depict $|t_{\mathrm{L}}|$ and $|t_{\mathrm{R}}|$. $|t_{\mathrm{L}}|$
diverges at $\phi _{\alpha }=2\pi /3$, vanishes at $\phi _{\beta }=-2\pi /3$%
, and becomes unity at $\phi _{\alpha }=2\pi /3$ and $\phi _{\beta }=-2\pi /3
$, where the SS of the side-coupled structures coincide. Figure~\ref{figPT}%
(d) depicts the scattering coefficients at $|\phi _{\alpha }|=2\pi /3$ ($r_{%
\mathrm{L}}$ divergence), implied by the dash-dotted blue line [$|t_{\mathrm{%
L}}|$ in Fig.~\ref{figPT}(b)] and solid red line [$|t_{\mathrm{R}}|$ in Fig.~%
\ref{figPT}(c)]. At $\phi _{\alpha }+\phi _{\beta }=0$, the system is $%
\mathcal{PT}$-symmetric; the scattering coefficients converge when $\phi
_{\alpha }=-\phi _{\beta }=\pm 2\pi /3$. $\mathcal{PT}$ symmetry ensures
that the persistent wave emission from resonator $\alpha $ is directly
absorbed at resonator $\beta $ and forms a unity transmittivity. The
scattering coefficients satisfy $|r_{\mathrm{L}}|=|t_{\mathrm{L}}|=|t_{%
\mathrm{R}}|=1$, $|r_{\mathrm{R}}|=0$ and the SS vanishes. At $\phi _{\alpha
}=\phi _{\beta }=2\pi /3$, a persistent right-going wave emission for the
left incidence and a perfect absorption for the right incidence occur: $%
\left\vert r_{\mathrm{L}}\right\vert =1$, $\left\vert t_{\mathrm{L}%
}\right\vert \rightarrow \infty $,$\ $and $r_{\mathrm{R}}=t_{\mathrm{R}}=0$.
At $\phi _{\alpha }=\phi _{\beta }=-2\pi /3$, a persistent left-going wave
emission for the right incidence and a full reflection for the left
incidence occur~\cite{Supplementary}.

\emph{Simultaneous unidirectional lasing and perfect absorption}.---UPA
prevents the backward flow without affecting the forward propagation, which
is a versatile building block for light manipulation. Series combination of
several UPAs and ULs enables more intriguing asymmetric phenomena (Fig.~\ref%
{fig4}).

Connecting a right-UPA on the left side of the two side-coupled resonators
in the situation marked by the magenta circle in Fig.~\ref{figPT}, the
finite left reflection from the right two side-coupled resonators is
perfectly absorbed; consequently, a reflectionless left incident
unidirectional lasing and right incident perfect absorption [Figs.~\ref{fig4}%
(a) and~\ref{fig4}(b)] is achieved in the configuration of Fig.~\ref{fig4}%
(c). If all the synthetic magnetic fluxes in Fig.~\ref{fig4}(c) are
opposite, which correspond to the configuration experienced by the CW mode
[Fig.~\ref{fig4}(f)], the dynamics switch between the left and right
incidences in contrast to the CCW mode [Figs.~\ref{fig4}(d) and~\ref{fig4}%
(e)]. Unidirectional lasing of different modes are toward opposite
directions. Simultaneously exciting the CCW and CW modes in the left (right)
side, the CCW (CW) mode induces a unidirectional lasing toward the right
(left) side and the CW (CCW) mode is perfectly absorbed.

\begin{figure}[t]
\centering\includegraphics[ bb=20 0 470 240, width=8.8 cm, clip]{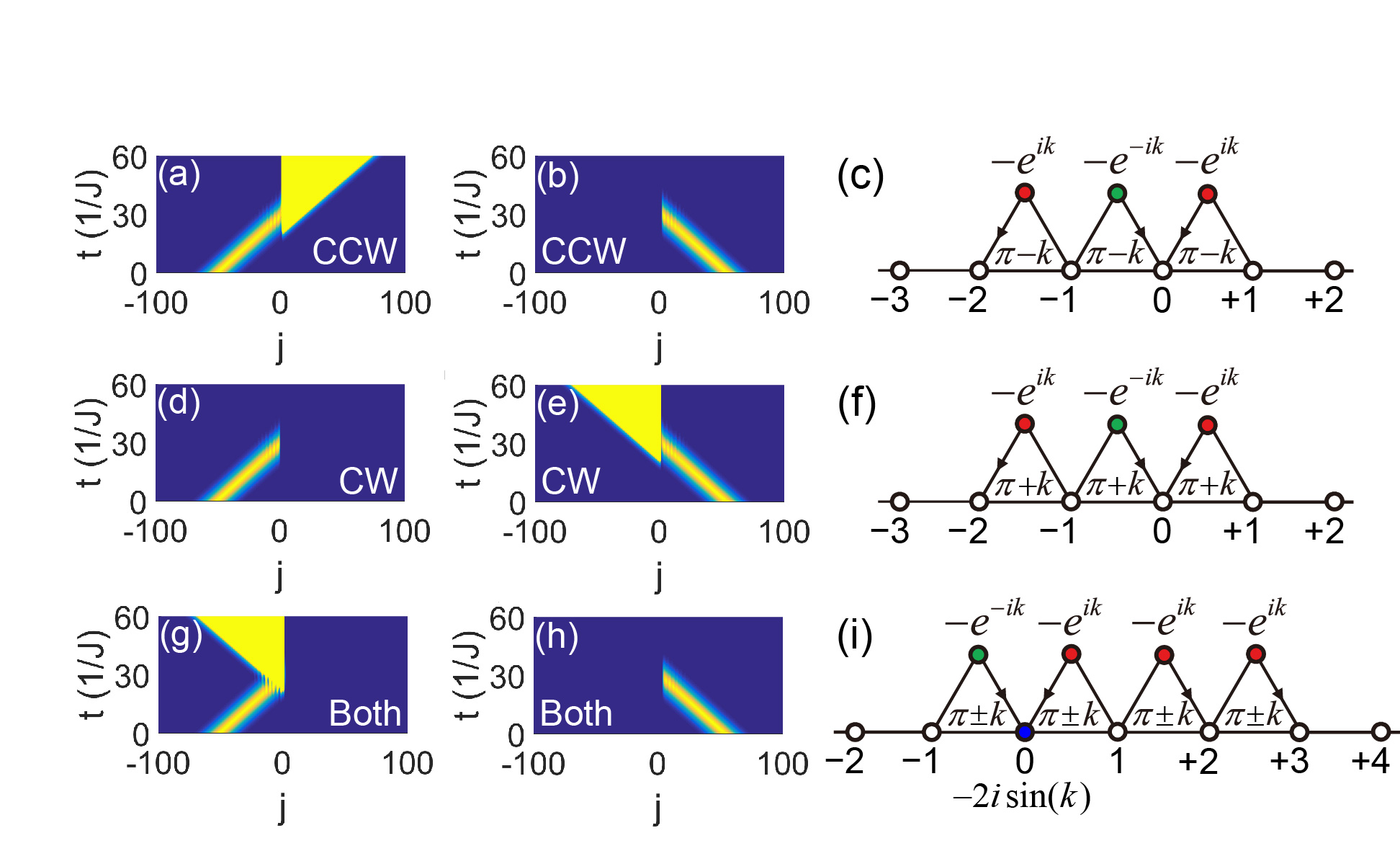}
\caption{Snapshots of the wave intensities and schematics of the equivalent systems for (a-f) transmissionless and (g-i) reflectionless unidirectional lasing and
perfect absorption for a Gaussian wave packet of $\protect\sigma =0.1$, $k_{\mathrm{c}}=\protect\pi /3$. The arrows
indicate the phase directions with values inside, $k=\protect\pi /3$.} \label%
{fig4}
\end{figure}

Connecting two right-UPAs on the right side of a single-direction lasing
[Figs.~\ref{fig2}(c) and~\ref{fig2}(d)] to form the configuration shown in
Fig.~\ref{fig4}(i), the right incidence is perfectly absorbed before
inducing a unidirectional lasing, resulting in a transmissionless left
incident unidirectional lasing and right incident perfect absorption [Figs.~%
\ref{fig4}(g) and~\ref{fig4}(h)]. The CW mode experiences opposite synthetic
magnetic fluxes, a left incident bidirectional lasing and right incident
perfect absorption is realized by the left two side-coupled resonators~\cite%
{Supplementary} and the right two right-UPAs change into left-UPAs, which
perfectly absorbing the right-going lasing. Notably, both the CCW and CW
modes possess identical dynamical phenomena.

\emph{Discussion}.---In summary, the bidirectional reflectionless as a
desirable feature of UDI allows scalable combination of unidirectional
dynamics, properly assembling distinct UDIs enriches the unidirectionality.
The incident direction independent wave propagation is proposed, including
the unidirectional lasing toward single definite direction and the one-way
propagation. The simultaneous unidirectional lasing and perfect absorption
is allowed. Our findings are applicable in optical waveguides~\cite{LonghiOL}%
.

The imperfections in the resonator as defects and surface roughness result
in backscattering~\cite{TJK,Borselli,Morichetti}, which induces a mode
coupling and mixes the CCW and CW modes~\cite{TJK}. The backscattering results in mode interchanging and is
unfavourable for the desirable unidirectional functionalities~\cite{LJinPRA2011}. Directional couplers can be used to reduce the influence of
backscattering~\cite{HafeziNatPhoton}. As shown in the Supplementary
Material~\cite{Supplementary}, UPA absorbs one mode without affecting the
other mode, which helps preventing the unwanted backscattering induced mode
accumulation in one side; moreover, the performances of UPA,
single-direction lasing, and transmissionless unidirectional lasing and
perfect absorption remain good at weak backscattering. Alternatively, the
backscattering is a useful resource for optical sensing~\cite%
{Wiersig,Wiersig2016,EP2Sensing}.

Synthetic magnetic flux has been realized in quantum regime~\cite%
{NP13,Goldman,PRA93}, it would be interesting to investigate the
unidirectionality in quantum dots, cold-atoms, or trapped-ions in the
frameworks of non-Hermitian physics and chiral quantum optics~\cite%
{TR,TRPRA,PL,PRA93,PZ}. Our findings open up new directions for designing
novel lasers and optical control devices including but not limited to laser,
absorber, rectifier, isolator, and modulator in a variety of areas in optics
and beyond.

\begin{acknowledgements}
This work was supported by NSFC (Grant No. 11605094) and Tianjin Natural
Science Foundation (Grant No. 16JCYBJC40800).
\end{acknowledgements}

\clearpage
\begin{widetext}
\section*{Supplementary Material for ``Incident Direction Independent Wave
Propagation and Unidirectional Lasing"}

\begin{center}
L. Jin* and Z. Song\\[2pt]
\textit{School of Physics, Nankai University, Tianjin 300071, China}
\end{center}

\subsection{Reciprocity of scattering coefficients}

The reciprocity of transmission or reflection ($t_{\mathrm{L}}=t_{\mathrm{R}%
} $ or $r_{\mathrm{L}}=r_{\mathrm{R}}$) is protected by the system symmetry~%
\cite{MugaPR,PTScattAnnPhys}. For a scattering system of Hamiltonian $H$,
under the notations of the parity operator $\mathcal{P}$ and the
time-reversal operator $\mathcal{T}$, the system is parity symmetric when $%
\mathcal{P}H\mathcal{P}^{-1}=H$, the system is time-reversal symmetric when $%
\mathcal{T}H\mathcal{T}^{-1}=H$, and the system is parity-time ($\mathcal{PT}
$) symmetric when $\mathcal{(PT)}H(\mathcal{PT})^{-1}=H$. The parity
symmetry leads to both reciprocal transmission and reflection ($t_{\mathrm{L}%
}=t_{\mathrm{R}}$ and $r_{\mathrm{L}}=r_{\mathrm{R}}$). For a system that
satisfies $\mathcal{T}H\mathcal{T}^{-1}=H^{\dagger }$, the system
transmission is reciprocal ($t_{\mathrm{L}}=t_{\mathrm{R}}$); in contrast,
the $\mathcal{PT}$-symmetry leads to symmetric transmission amplitude ($%
\left\vert t_{\mathrm{L}}\right\vert =\left\vert t_{\mathrm{R}}\right\vert $%
).

For the Hermitian system $H=H^{\dagger }$, the time-reversal symmetry
indicates $\mathcal{T}H\mathcal{T}^{-1}=H=H^{\dagger }$, thus the system
transmission is reciprocal ($t_{\mathrm{L}}=t_{\mathrm{R}}$). For the
non-Hermitian system, $\mathcal{T}H\mathcal{T}^{-1}=H^{\dagger }$\ is
satisfied in the $\mathcal{PT}$-symmetric systems that possessing gain/loss
and the Hermitian symmetric couplings, where reciprocal transmissions ($t_{%
\mathrm{L}}=t_{\mathrm{R}}$) are observed~\cite{Chong11,LFeng,USS}; other $%
\mathcal{PT}$-symmetric non-Hermitian systems may satisfy $\mathcal{T}H%
\mathcal{T}^{-1}=H^{\dagger }$. In this Letter, the synthetic magnetic flux
induced by the asymmetric coupling for unidirectional destructive
interference as shown in Fig. 1(a) (Fig.~\ref{figUPA})
breaks the parity symmetry and leads to $\mathcal{T}H\mathcal{T}^{-1}\neq
H^{\dagger }$, where we notice a nonreciprocal transmission ($t_{\mathrm{L}%
}\neq t_{\mathrm{R}}$) from the scattering matrix $S$ even if the system is
Hermitian with real $V_{\alpha }$.

\subsection{Scattering coefficients of resonator chain with one side-coupled
resonator}

The unidirectional destructive interference is valid under unequal coupling
strengths between the side-coupling and the uniform coupling of the
resonator chain. We consider resonator $\alpha $ side-coupled to the uniform
resonator chain as schematically illustrated in Figs.~\ref%
{figUPA}(a) and~\ref{figUPA}(b), and resonator $0$ is on resonance with
other resonators in the chain. The resonator chain has a uniform coupling $J$%
, the two side-couplings are $g$ and $ge^{\pm i\phi _{\alpha }}$. The
special case of $g=J$ is presented in the Letter, a general case is
considered here.

In the coupled-mode theory~\cite{Haus,Joannopoulos}, the equation of motion
for the resonator $j\neq -1$, $0$, and $\alpha $ in the system illustrated
in Fig.~\ref{figUPA}(a) is%
\begin{equation}
i\frac{\mathrm{d}\psi _{j}}{\mathrm{d}t}=\omega _{\mathrm{c}}\psi _{j}-J\psi
_{j-1}-J\psi _{j+1},  \label{1}
\end{equation}%
and the equations of motion for all the other resonators $j=-1$, $0$, and $%
\alpha $ are
\begin{eqnarray}
&&i\frac{\mathrm{d}\psi _{-1}}{\mathrm{d}t}=\omega _{\mathrm{c}}\psi
_{-1}-J\psi _{-2}-J\psi _{0}-g\psi _{\alpha },  \label{2} \\
&&i\frac{\mathrm{d}\psi _{\alpha }}{\mathrm{d}t}=\left( \omega _{\mathrm{c}%
}+V_{\alpha }\right) \psi _{\alpha }-g\psi _{-1}-ge^{-i\phi _{\alpha }}\psi
_{0},  \label{3} \\
&&i\frac{\mathrm{d}\psi _{0}}{\mathrm{d}t}=\omega _{\mathrm{c}}\psi
_{0}-J\psi _{-1}-J\psi _{1}-ge^{i\phi _{\alpha }}\psi _{\alpha },  \label{4}
\end{eqnarray}%
where $\omega _{\mathrm{c}}=2\pi c/\lambda $ is the resonator frequency with
$\lambda $ the wavelength and $c$ the light velocity in vacuum. For the
dimensionless Bloch wave vector $k$, the dispersion relation supported by
the resonator chain is $\omega =\omega _{\mathrm{c}}-2J\cos k$ and the field
amplitude is $\psi _{j}=f_{j}e^{-i\omega t}$. We obtain the equations of
motion for the resonators $j=-1$, $0$ and $\alpha $ at steady-state as
\begin{eqnarray}
&&\omega f_{-1}=\omega _{\mathrm{c}}f_{-1}-Jf_{-2}-Jf_{0}-gf_{\alpha },
\label{5} \\
&&\omega f_{\alpha }=\left( \omega _{\mathrm{c}}+V_{\alpha }\right)
f_{\alpha }-gf_{-1}-ge^{-i\phi _{\alpha }}f_{0},  \label{6} \\
&&\omega f_{0}=\omega _{\mathrm{c}}f_{0}-Jf_{-1}-Jf_{1}-ge^{i\phi _{\alpha
}}f_{\alpha }.  \label{7}
\end{eqnarray}

The steady-state wave functions are in the form of $f_{j}=Ae^{ikj}+Be^{-ikj}$
($j<0$) and $f_{j}=Ce^{ikj}+De^{-ikj}$ ($j\geqslant 0$) for the resonator $j$
in the elastic scattering process. Thus, the wave functions of resonators $%
j=0$, $\pm 1$, $\pm 2$ are $f_{-2}=Ae^{-2ik}+Be^{2ik}$, $%
f_{-1}=Ae^{-ik}+Be^{ik}$, $f_{0}=C+D$, $f_{1}=Ce^{ik}+De^{-ik}$, and $%
f_{2}=Ce^{2ik}+De^{-2ik}$. Substituting the wave functions and the
dispersion relation into Eqs.~(\ref{5}-\ref{7}), the coefficients in the
steady-state wave functions satisfy%
\begin{eqnarray}
\left( -e^{-ik}-e^{ik}\right) \left( Ae^{-ik}+Be^{ik}\right) &=&-\left(
Ae^{-2ik}+Be^{2ik}\right) -\left( C+D\right) -\frac{g}{J}f_{\alpha }, \\
\left( -e^{-ik}-e^{ik}-\frac{V_{\alpha }}{J}\right) f_{\alpha } &=&-\frac{g}{%
J}\left( Ae^{-ik}+Be^{ik}\right) -\frac{g}{J}e^{-i\phi _{\alpha }}\left(
C+D\right) , \\
\left( -e^{-ik}-e^{ik}\right) \left( C+D\right) &=&-\left(
Ae^{-ik}+Be^{ik}\right) -\left( Ce^{ik}+De^{-ik}\right) -\frac{g}{J}e^{i\phi
_{\alpha }}f_{\alpha }.
\end{eqnarray}%
After eliminating $f_{\alpha }$, the coefficients satisfy%
\begin{eqnarray}
\left( e^{-ik}+e^{ik}+\frac{V_{\alpha }}{J}\right) \left( A+B-C-D\right) -%
\frac{g^{2}}{J^{2}}\left( Ae^{-ik}+Be^{ik}\right) -\frac{g^{2}}{J^{2}}%
e^{-i\phi _{\alpha }}\left( C+D\right) &=&0,  \label{S1} \\
\left( Ae^{-ik}+Be^{ik}\right) +e^{i\phi _{\alpha }}\left( A+B-C-D\right)
-\left( Ce^{-ik}+De^{ik}\right) &=&0.  \label{S2}
\end{eqnarray}

\begin{figure}[thb]
\includegraphics[ bb=0 0 540 390, width=14.8 cm, clip]{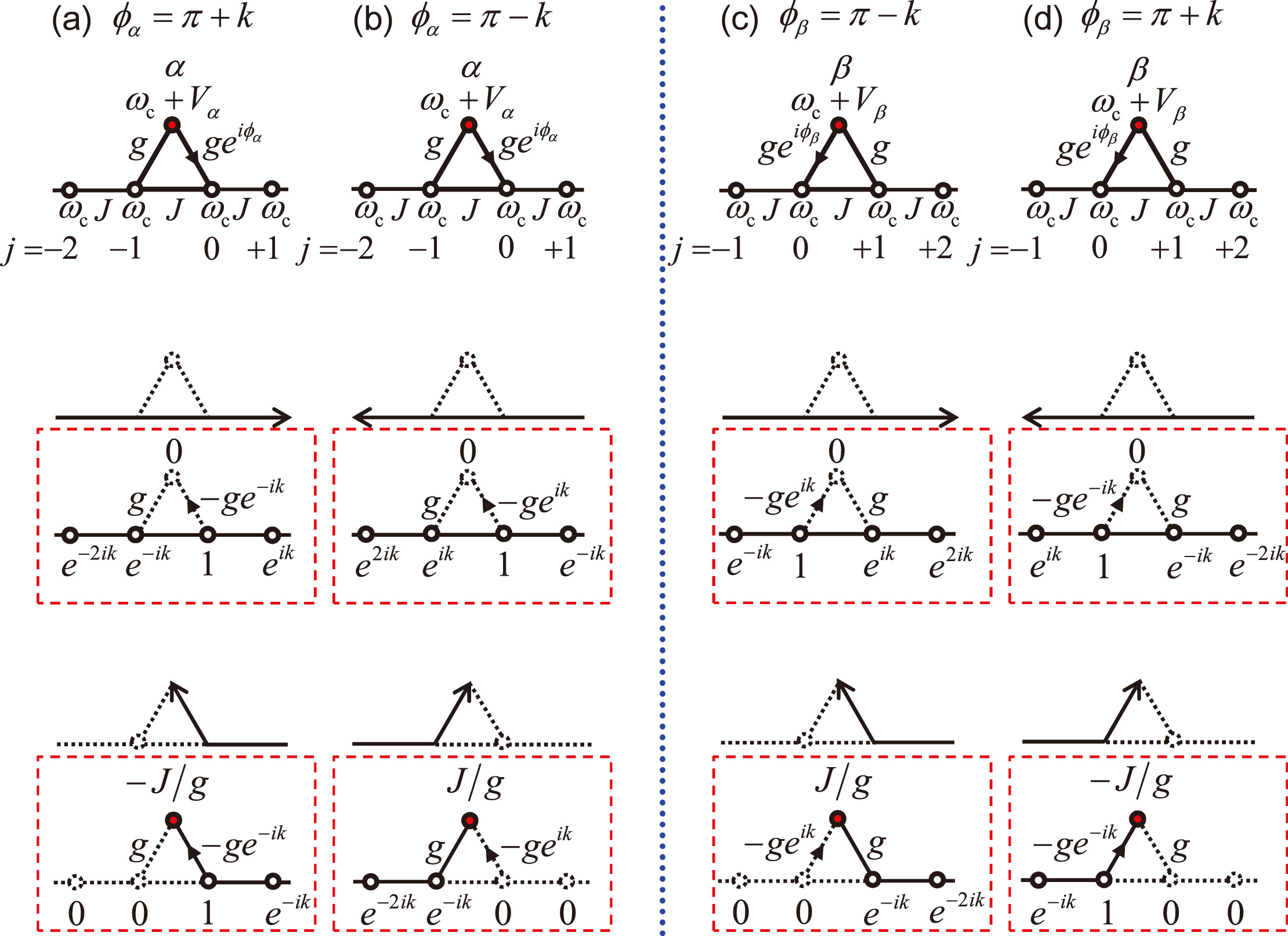}
\caption{Unidirectional perfect absorption. (a, c) Right-UPA. (b, d)
Left-UPA. The upper panels illustrate side-coupled configurations with
distinct synthetic magnetic fluxes. The values of $\protect\phi_{\protect\alpha}$ and $\protect\phi_{\protect\beta}$
for the unidirectional perfect absorption are marked on the top of the
schematics. $j$ is the resonator index, $k$ is the
incident wave vector, and $V_{\protect\alpha }=(g^{2}/J)e^{-ik}-2J\cos k$.
The central panels illustrate the destructive interference from one incident
direction and the lower panels illustrate the perfect absorption from the
opposite incident direction. The wave functions and side-couplings are shown
inside the red rectangles, the phase directions in the side-couplings are
indicated. The dotted lines indicate the equivalently decoupled part.} \label%
{figUPA}
\end{figure}

As schematically illustrated in Fig.~\ref{figUPA}(a), the
synthetic magnetic flux is $\phi _{\alpha }=\pi +k$~\cite{FLUX}. Thus,
equations~(\ref{S1},~\ref{S2}) reduce to
\begin{eqnarray}
\left( e^{-ik}+e^{ik}+\frac{V_{\alpha }}{J}\right) \left( A+B\right) -\frac{%
g^{2}}{J^{2}}\left( Ae^{-ik}+Be^{ik}\right) -\left( e^{-ik}+e^{ik}+\frac{%
V_{\alpha }}{J}-\frac{g^{2}}{J^{2}}e^{-ik}\right) \left( C+D\right) &=&0,
\label{S3} \\
\left( A-C\right) \left( e^{-ik}-e^{ik}\right) &=&0.  \label{S4}
\end{eqnarray}%
For the left incidence ($D=0$); substituting $D=0$ into Eqs.~(\ref{S3},~\ref%
{S4}), we obtain $B=0$ and $C=A$. Therefore, $r_{\mathrm{L}}=B/A=0$ and $t_{%
\mathrm{L}}=C/A=1$. For the right incidence ($A=0$); substituting $A=0$ into
Eqs.~(\ref{S1},~\ref{S2}), we obtain $C=0$ and
\begin{equation}
\left( e^{-ik}+\frac{V_{\alpha }}{J}+e^{ik}-\frac{g^{2}}{J^{2}}e^{ik}\right)
B=\left( e^{ik}+\frac{V_{\alpha }}{J}+e^{-ik}-\frac{g^{2}}{J^{2}}%
e^{-ik}\right) D.
\end{equation}%
Therefore, $r_{\mathrm{R}}=C/D=0$ and
\begin{equation}
t_{\mathrm{R}}=\frac{B}{D}=\frac{J^{2}e^{ik}+JV_{\alpha }+\left(
J^{2}-g^{2}\right) e^{-ik}}{J^{2}e^{-ik}+JV_{\alpha }+\left(
J^{2}-g^{2}\right) e^{ik}}.
\end{equation}%
The scattering is reflectionless for the incidences from both sides. At $%
g=J=1$, the scattering matrix~\cite{Jalas} reduces to
\begin{equation}
S=\left(
\begin{array}{cc}
r_{\mathrm{L}} & t_{\mathrm{R}} \\
t_{\mathrm{L}} & r_{\mathrm{R}}%
\end{array}%
\right) =\left(
\begin{array}{cc}
0 & \frac{J^{2}e^{ik}+JV_{\alpha }+\left( J^{2}-g^{2}\right) e^{-ik}}{%
J^{2}e^{-ik}+JV_{\alpha }+\left( J^{2}-g^{2}\right) e^{ik}} \\
1 & 0%
\end{array}%
\right) =\left(
\begin{array}{cc}
0 & \frac{e^{ik}+V_{\alpha }}{e^{-ik}+V_{\alpha }} \\
1 & 0%
\end{array}%
\right) .  \label{S}
\end{equation}

In Fig.~\ref{figUPA}, the steady-state wave functions are
listed for the destructive interference in the middle panels and for the
perfect absorption in the lower panels inside the red rectangles. The
direction dependent asymmetric couplings are shown with the phase directions
indicated by the arrows. The validity of steady-state wave functions can be
directly checked in the equations of motion. For example, in the middle
panel of Fig.~\ref{figUPA}(a), the contributions from
resonators $-1$ and $0$ (wave functions multiple the corresponding
couplings) cancel each other at resonator $\alpha $, leading to a completely
destructive interference; thus, resonator $\alpha $ is equivalently
isolated; in the lower panel of Fig.~\ref{figUPA}(a), the
left-going wave does not pass through resonator $\alpha $, the nonvanishing
wave function of resonator $\alpha $ results in a wave absorption and the
contributions from resonators $\alpha $ and $0$ (wave functions multiple the
corresponding couplings) cancel at resonator $-1$, then the left half ($j<0$%
) of the resonator chain is equivalently decoupled.

As schematically illustrated in Fig.~\ref{figUPA}(b), the
synthetic magnetic flux is $\phi _{\alpha }=\pi -k$. Thus, equations~(\ref%
{S1},~\ref{S2}) reduce to%
\begin{eqnarray}
\left( e^{-ik}+e^{ik}+\frac{V_{\alpha }}{J}\right) \left( A+B\right) -\frac{%
g^{2}}{J^{2}}\left( Ae^{-ik}+Be^{ik}\right) -\left( e^{-ik}+e^{ik}+\frac{%
V_{\alpha }}{J}-\frac{g^{2}}{J^{2}}e^{ik}\right) \left( C+D\right) &=&0,
\label{S5} \\
\left( B-D\right) \left( e^{ik}-e^{-ik}\right) &=&0.  \label{S6}
\end{eqnarray}%
Similarly, the reflection and transmission coefficients are calculated from
Eqs.~(\ref{S5},~\ref{S6}). For the left incidence ($D=0$), we obtain $B=0$
and%
\begin{equation}
\left( e^{-ik}+e^{ik}+\frac{V_{\alpha }}{J}-\frac{g^{2}}{J^{2}}e^{ik}\right)
C=\left( e^{-ik}+e^{ik}+\frac{V_{\alpha }}{J}-\frac{g^{2}}{J^{2}}%
e^{-ik}\right) A.
\end{equation}%
Therefore, $r_{\mathrm{L}}=B/A=0$ and
\begin{equation}
t_{\mathrm{L}}=\frac{C}{A}=\frac{J^{2}e^{ik}+JV_{\alpha }+\left(
J^{2}-g^{2}\right) e^{-ik}}{J^{2}e^{-ik}+JV_{\alpha }+\left(
J^{2}-g^{2}\right) e^{ik}}.
\end{equation}%
For the right incidence ($A=0$); we have $C=0$ and $B=D$. Therefore, $r_{%
\mathrm{R}}=C/D=0$ and $t_{\mathrm{R}}=B/D=1$. At $g=J=1$, the scattering
matrix reduces to%
\begin{equation}
S=\left(
\begin{array}{cc}
r_{\mathrm{L}} & t_{\mathrm{R}} \\
t_{\mathrm{L}} & r_{\mathrm{R}}%
\end{array}%
\right) =\left(
\begin{array}{cc}
0 & 1 \\
\frac{J^{2}e^{ik}+JV_{\alpha }+\left( J^{2}-g^{2}\right) e^{-ik}}{%
J^{2}e^{-ik}+JV_{\alpha }+\left( J^{2}-g^{2}\right) e^{ik}} & 0%
\end{array}%
\right) =\left(
\begin{array}{cc}
0 & 1 \\
\frac{e^{ik}+V_{\alpha }}{e^{-ik}+V_{\alpha }} & 0%
\end{array}%
\right) .
\end{equation}

Figures~\ref{figUPA}(a) and~\ref{figUPA}(b) are the situations
that the asymmetric coupling is on the right side of the triangular
structure. Figs.~\ref{figUPA}(c) and~\ref{figUPA}(d) are the
situations that the asymmetric coupling is on the left side of the
triangular structure. The structures shown in Figs.~\ref%
{figUPA}(c) and~\ref{figUPA}(d) are the left-right mirror reflection
(indicated by the blue dotted line) of that shown in Figs.~%
\ref{figUPA}(b) and~\ref{figUPA}(a), respectively. This can be recognized
after substituting the resonator indexes $\beta \rightarrow \alpha $ and $%
j\rightarrow -j$.

\subsection{Scattering coefficients of resonator chain with two side-coupled
resonators}

We consider equal coupling $g=J=1$ in the Letter. The equations of motion
[Eqs. (2, 3, 7, 8, 9) in the Letter] at steady-state reduce to%
\begin{eqnarray}
&&\omega f_{-1}=\omega _{\mathrm{c}}f_{-1}-f_{-2}-f_{0}-f_{\alpha },
\label{Two1} \\
&&\omega f_{\alpha }=\left( \omega _{\mathrm{c}}+V_{\alpha }\right)
f_{\alpha }-f_{-1}-e^{-i\phi _{\alpha }}f_{0},  \label{Two2} \\
&&\omega f_{0}=\left( \omega _{\mathrm{c}}+V_{0}\right)
f_{0}-f_{-1}-f_{1}-e^{i\phi _{\alpha }}f_{\alpha }-e^{i\phi _{\beta
}}f_{\beta },  \label{Two3} \\
&&\omega f_{\beta }=\left( \omega _{\mathrm{c}}+V_{\beta }\right) f_{\beta
}-f_{1}-e^{-i\phi _{\beta }}f_{0},  \label{Two4} \\
&&\omega f_{1}=\omega _{\mathrm{c}}f_{1}-f_{0}-f_{2}-f_{\beta }.
\label{Two5}
\end{eqnarray}%
The steady-state wave functions are in form of $f_{j}=Ae^{ikj}+Be^{-ikj}$ ($%
j<0$) and $f_{j}=Ce^{ikj}+De^{-ikj}$ ($j>0$) for the resonator $j$.
Substituting the wave functions $f_{-2}=Ae^{-2ik}+Be^{2ik}$, $%
f_{-1}=Ae^{-ik}+Be^{ik}$, $f_{1}=Ce^{ik}+De^{-ik}$, $%
f_{2}=Ce^{2ik}+De^{-2ik} $, and the dispersion relation $\omega =\omega _{%
\mathrm{c}}-2\cos k$ into Eqs.$~$(\ref{Two1}-\ref{Two5}), we obtain
\begin{eqnarray}
\left( -e^{-ik}-e^{ik}\right) \left( Ae^{-ik}+Be^{ik}\right) &=&-\left(
Ae^{-2ik}+Be^{2ik}\right) -f_{0}-f_{\alpha }, \\
\left( -e^{-ik}-e^{ik}-V_{\alpha }\right) f_{\alpha } &=&-\left(
Ae^{-ik}+Be^{ik}\right) -e^{-i\phi _{\alpha }}f_{0}, \\
\left( -e^{-ik}-e^{ik}-V_{0}\right) f_{0} &=&-\left( Ae^{-ik}+Be^{ik}\right)
-\left( Ce^{ik}+De^{-ik}\right) -e^{i\phi _{\alpha }}f_{\alpha }-e^{i\phi
_{\beta }}f_{\beta }, \\
\left( -e^{-ik}-e^{ik}-V_{\beta }\right) f_{\beta } &=&-\left(
Ce^{ik}+De^{-ik}\right) -e^{-i\phi _{\beta }}f_{0}, \\
\left( -e^{-ik}-e^{ik}\right) \left( Ce^{ik}+De^{-ik}\right)
&=&-f_{0}-\left( Ce^{2ik}+De^{-2ik}\right) -f_{\beta }.
\end{eqnarray}%
After simplification, we have%
\begin{eqnarray}
f_{\alpha } &=&A+B-f_{0}, \\
\left( e^{ik}+V_{\alpha }\right) A+\left( e^{-ik}+V_{\alpha }\right) B
&=&\left( e^{-ik}+e^{ik}+V_{\alpha }+e^{-i\phi _{\alpha }}\right) f_{0}, \\
\left( e^{-ik}+e^{i\phi _{\alpha }}\right) A+\left( e^{ik}+e^{i\phi _{\alpha
}}\right) B+\left( e^{ik}+e^{i\phi _{\beta }}\right) C+\left(
e^{-ik}+e^{i\phi _{\beta }}\right) D &=&\left( e^{-ik}+e^{ik}+V_{0}+e^{i\phi
_{\alpha }}+e^{i\phi _{\beta }}\right) f_{0}, \\
\left( e^{-ik}+V_{\beta }\right) C+\left( e^{ik}+V_{\beta }\right) D
&=&\left( e^{-ik}+e^{ik}+V_{\beta }+e^{-i\phi _{\beta }}\right) f_{0}, \\
f_{\beta } &=&C+D-f_{0}.
\end{eqnarray}%
For the left incidence ($D=0$), we obtain the left reflection coefficient $%
r_{\mathrm{L}}=B/A$ and the left transmission coefficient $t_{\mathrm{L}%
}=C/A $,%
\begin{equation}
\begin{array}{l}
r_{\mathrm{L}}=\frac{\left( e^{ik}+V_{\alpha }\right) \left[ \left( e^{i\phi
_{\alpha }}+e^{-ik}+V_{0}\right) \left( e^{-ik}+V_{\beta }\right) -\left(
e^{ik}+e^{-i\phi _{\beta }}\right) \left( e^{i\phi _{\beta }}+e^{ik}\right) %
\right] -\left( e^{-ik}+e^{ik}+V_{\alpha }+e^{-i\phi _{\alpha }}\right)
\left( e^{-ik}+V_{\beta }\right) \left( e^{i\phi _{\alpha }}+e^{-ik}\right)
}{\left( e^{-ik}+V_{\alpha }\right) \left[ \left( e^{ik}+e^{-i\phi _{\beta
}}\right) \left( e^{i\phi _{\beta }}+e^{ik}\right) -\left( e^{i\phi _{\alpha
}}+e^{-ik}+V_{0}\right) \left( e^{-ik}+V_{\beta }\right) \right] +\left(
e^{-ik}+e^{ik}+V_{\alpha }+e^{-i\phi _{\alpha }}\right) \left(
e^{-ik}+V_{\beta }\right) \left( e^{i\phi _{\alpha }}+e^{ik}\right) }, \\
t_{\mathrm{L}}=\frac{\left( e^{-ik}+V_{\beta }+e^{ik}+e^{-i\phi _{\beta
}}\right) \left[ \left( e^{ik}+V_{\alpha }\right) \left( e^{i\phi _{\alpha
}}+e^{ik}\right) -\left( e^{-ik}+V_{\alpha }\right) \left( e^{i\phi _{\alpha
}}+e^{-ik}\right) \right] }{\left( e^{-ik}+V_{\alpha }\right) \left[ \left(
e^{ik}+e^{-i\phi _{\beta }}\right) \left( e^{i\phi _{\beta }}+e^{ik}\right)
-\left( e^{-ik}+V_{0}+e^{i\phi _{\alpha }}\right) \left( e^{-ik}+V_{\beta
}\right) \right] +\left( e^{-ik}+e^{ik}+V_{\alpha }+e^{-i\phi _{\alpha
}}\right) \left( e^{-ik}+V_{\beta }\right) \left( e^{i\phi _{\alpha
}}+e^{ik}\right) }.%
\end{array}
\label{L}
\end{equation}%
For the right incidence ($A=0$), we obtain the right reflection coefficient $%
r_{\mathrm{R}}=C/D$ and the right transmission coefficient $t_{\mathrm{R}%
}=B/D$,%
\begin{equation}
\begin{array}{l}
r_{\mathrm{R}}=\frac{\left( e^{ik}+V_{\beta }\right) \left[ \left( e^{i\phi
_{\beta }}+e^{-ik}+V_{0}\right) \left( e^{-ik}+V_{\alpha }\right) -\left(
e^{ik}+e^{-i\phi _{\alpha }}\right) \left( e^{ik}+e^{i\phi _{\alpha
}}\right) \right] -\left( e^{-ik}+e^{ik}+V_{\beta }+e^{-i\phi _{\beta
}}\right) \left( e^{-ik}+V_{\alpha }\right) \left( e^{-ik}+e^{i\phi _{\beta
}}\right) }{\left( e^{-ik}+V_{\beta }\right) \left[ \left( e^{ik}+e^{-i\phi
_{\alpha }}\right) \left( e^{ik}+e^{i\phi _{\alpha }}\right) -\left(
e^{-ik}+V_{\alpha }\right) \left( e^{i\phi _{\beta }}+e^{-ik}+V_{0}\right) %
\right] +\left( e^{-ik}+e^{ik}+V_{\beta }+e^{-i\phi _{\beta }}\right) \left(
e^{-ik}+V_{\alpha }\right) \left( e^{ik}+e^{i\phi _{\beta }}\right) }, \\
t_{\mathrm{R}}=\frac{\left( e^{-ik}+e^{ik}+V_{\alpha }+e^{-i\phi _{\alpha
}}\right) \left[ \left( e^{ik}+e^{i\phi _{\beta }}\right) \left(
e^{ik}+V_{\beta }\right) -\left( e^{-ik}+e^{i\phi _{\beta }}\right) \left(
e^{-ik}+V_{\beta }\right) \right] }{\left( e^{-ik}+V_{\beta }\right) \left[
\left( e^{ik}+e^{i\phi _{\alpha }}\right) \left( e^{ik}+e^{-i\phi _{\alpha
}}\right) -\left( e^{-ik}+V_{\alpha }\right) \left( e^{i\phi _{\beta
}}+e^{-ik}+V_{0}\right) \right] +\left( e^{-ik}+e^{ik}+V_{\beta }+e^{-i\phi
_{\beta }}\right) \left( e^{-ik}+V_{\alpha }\right) \left( e^{ik}+e^{i\phi
_{\beta }}\right) }.%
\end{array}
\label{R}
\end{equation}

The interference is clear when the system is not at spectral singularities.
In the case of $e^{ik}+e^{-i\phi _{\beta }}=0$, $r_{\mathrm{L}}$ and $t_{%
\mathrm{L}}$ are irrelevant to $V_{\beta }$, the side-coupled resonator $%
\beta $ is isolated due to the destructive interference for the left
incidence; thus, the scattering is only affected by $V_{\alpha }$ and $V_{0}$%
. On the contrary, when $e^{ik}+e^{-i\phi _{\alpha }}=0$, resonator $\alpha $
is isolated for the right incidence and $r_{\mathrm{R}}$ and $t_{\mathrm{R}}$
are irrelevant to $V_{\alpha }$. When $e^{ik}+e^{i\phi _{\beta }}=0$, $r_{%
\mathrm{L}}$ and $t_{\mathrm{R}}$ are $V_{\beta }$ irrelevant. This
indicates that $V_{\beta }$ does not affect the left-going waves (for
incidence from either side). Similarly, when $e^{ik}+e^{i\phi _{\alpha }}=0$%
, $t_{\mathrm{L}}$ and $r_{\mathrm{R}}$ are $V_{\alpha }$ irrelevant; $%
V_{\alpha }$ does not affect the right-going waves (for incidence from
either side). Different matches of the synthetic magnetic fluxes result in
different influences of the side-coupled resonators.

In the combination of $e^{ik}+e^{i\phi _{\alpha }}=0$ and $e^{ik}+e^{i\phi
_{\beta }}=0$, e.g., $\phi _{\alpha }=\phi _{\beta }=\pi +k$, the left
reflection and right transmission coefficients (the left-going waves) are $%
V_{\beta }$ irrelevant; the left transmission and right reflection
coefficients (the right-going waves) are $V_{\alpha }$ irrelevant. The
reflection and transmission coefficients reduce to%
\begin{eqnarray}
r_{\mathrm{L}} &=&\frac{-\left( e^{ik}+V_{\alpha }\right) V_{0}}{\left(
e^{-ik}+V_{\alpha }\right) \left( e^{i\phi _{\alpha }}+e^{-ik}+V_{0}\right) }%
,t_{\mathrm{L}}=\frac{\left( e^{ik}+V_{\beta }\right) \left( e^{i\phi
_{\alpha }}+e^{-ik}\right) }{\left( e^{-ik}+V_{0}+e^{i\phi _{\alpha
}}\right) \left( e^{-ik}+V_{\beta }\right) }, \\
r_{\mathrm{R}} &=&\frac{-\left( e^{ik}+V_{\beta }\right) V_{0}}{\left(
e^{-ik}+V_{\beta }\right) \left( e^{i\phi _{\beta }}+e^{-ik}+V_{0}\right) }%
,t_{\mathrm{R}}=\frac{\left( e^{ik}+V_{\alpha }\right) \left(
e^{-ik}+e^{i\phi _{\beta }}\right) }{\left( e^{-ik}+V_{\alpha }\right)
\left( e^{i\phi _{\beta }}+e^{-ik}+V_{0}\right) }.
\end{eqnarray}%
One-way propagation occurs at $V_{\alpha }=-\cos k+3^{\pm 1}i\sin k$, $%
V_{\beta }=-e^{ik}$, and $V_{0}=-2i\sin k$. Resonator $\alpha $ tunes the
amplitude of the left-going propagating wave; resonator $\beta $ absorbs the
right-going propagating wave. The wave is either completely left reflected
or completely right transmitted, i.e., $\left\vert r_{\mathrm{L}}\right\vert
=\left\vert t_{\mathrm{R}}\right\vert =1$ and $t_{\mathrm{L}}=r_{\mathrm{R}%
}=0$. Notably, the one-way propagation can be realized in another way by
assembling one more side-coupled resonator. For example, assembling the
unidirectional perfect absorption structure of left perfect absorption and
right resonant transmission [left-UPA in Fig.~\ref{figUPA}%
(b) or~\ref{figUPA}(d)] on the right side of the $\mathcal{PT}$-symmetric
two side-coupled structure in the situations of Fig.~\ref%
{figPTdynamics}(a) or~\ref{figPTdynamics}(d) in section C enables the
one-way propagation. A single-direction lasing occurs at $\phi _{\alpha
}=\phi _{\beta }=\pi +k$, $V_{\alpha }=-e^{-ik}$, $V_{0}=-2i\sin k$, and $%
V_{\beta }=-e^{ik}$.

In the combination of $e^{ik}+e^{-i\phi _{\alpha }}=0$ and $e^{ik}+e^{-i\phi
_{\beta }}=0$, e.g., $\phi _{\alpha }=\phi _{\beta }=\pi -k$, the left
(right) reflection and transmission coefficients are $V_{\beta (\alpha )}$
irrelevant. The reflection and transmission coefficients reduce to
\begin{eqnarray}
r_{\mathrm{L}} &=&\frac{\left( e^{ik}+V_{\alpha }\right) V_{0}}{\left(
e^{-ik}+V_{\alpha }\right) \left( e^{i\phi _{\alpha }}+e^{ik}-V_{0}\right) }%
,t_{\mathrm{L}}=\frac{\left( e^{ik}+V_{\alpha }\right) \left( e^{i\phi
_{\alpha }}+e^{ik}\right) }{\left( e^{-ik}+V_{\alpha }\right) \left(
e^{i\phi _{\alpha }}+e^{ik}-V_{0}\right) }, \\
r_{\mathrm{R}} &=&\frac{\left( e^{ik}+V_{\beta }\right) V_{0}}{\left(
e^{-ik}+V_{\beta }\right) \left( e^{ik}+e^{i\phi _{\beta }}-V_{0}\right) }%
,t_{\mathrm{R}}=\frac{\left( e^{ik}+V_{\beta }\right) \left( e^{ik}+e^{i\phi
_{\beta }}\right) }{\left( e^{-ik}+V_{\beta }\right) \left( e^{ik}+e^{i\phi
_{\beta }}-V_{0}\right) }.
\end{eqnarray}%
In practice, this case reveals the dynamics of the CW mode when the CCW mode
experienced the synthetic magnetic fluxes $\phi _{\alpha }=\phi _{\beta
}=\pi +k$ in the system. A simultaneous left incident bidirectional lasing
and right incident perfect absorption occurs at $\phi _{\alpha }=\phi
_{\beta }=\pi -k$, $V_{\alpha }=-e^{-ik}$, $V_{0}=-2i\sin k$, and $V_{\beta
}=-e^{ik}$.

\subsection{Scattering coefficients of $\mathcal{PT}$-symmetric resonator
chain with two side-coupled resonators}

We consider equal coupling $g=J=1$ in the Letter, the situation of $\mathcal{%
PT}$-symmetric resonator chain is discussed in this section. Resonator $0$
is on resonance with other resonators in the chain ($V_{0}=0$), the
side-coupled resonators are $\mathcal{PT}$-symmetric, $V_{\alpha }=-e^{-ik}$
and $V_{\beta }=-e^{ik}$. Substituting the wave functions $%
f_{-2}=Ae^{-2ik}+Be^{2ik}$, $f_{-1}=Ae^{-ik}+Be^{ik}$, $%
f_{1}=Ce^{ik}+De^{-ik}$ and $f_{2}=Ce^{2ik}+De^{-2ik}$, and the dispersion
relation $\omega =\omega _{\mathrm{c}}-2\cos k$ into the equations of motion
at steady-state [Eqs.$~$(\ref{Two1}-\ref{Two5})], we have%
\begin{eqnarray}
f_{\alpha } &=&A+B-f_{0}, \\
\left( e^{ik}+e^{-ik}+V_{\alpha }\right) f_{\alpha } &=&\left(
Ae^{-ik}+Be^{ik}\right) +e^{-i\phi _{\alpha }}f_{0}, \\
\left( e^{ik}+e^{-ik}\right) f_{0} &=&\left( Ae^{-ik}+Be^{ik}\right) +\left(
Ce^{ik}+De^{-ik}\right) +e^{i\phi _{\alpha }}f_{\alpha }+e^{i\phi _{\beta
}}f_{\beta }, \\
\left( e^{ik}+e^{-ik}+V_{\beta }\right) f_{\beta } &=&\left(
Ce^{ik}+De^{-ik}\right) +e^{-i\phi _{\beta }}f_{0}, \\
f_{\beta } &=&C+D-f_{0},
\end{eqnarray}%
after simplification, we obtain
\begin{eqnarray}
\left( e^{ik}+V_{\alpha }\right) A+\left( e^{-ik}+V_{\alpha }\right) B
&=&\left( e^{ik}+e^{-ik}+V_{\alpha }+e^{-i\phi _{\alpha }}\right) f_{0}, \\
\left( e^{ik}+e^{-ik}+e^{i\phi _{\alpha }}+e^{i\phi _{\beta }}\right) f_{0}
&=&\left( e^{-ik}+e^{i\phi _{\alpha }}\right) A+\left( e^{ik}+e^{i\phi
_{\alpha }}\right) B+\left( e^{ik}+e^{i\phi _{\beta }}\right) C+\left(
e^{-ik}+e^{i\phi _{\beta }}\right) D, \\
\left( e^{-ik}+V_{\beta }\right) C+\left( e^{ik}+V_{\beta }\right) D
&=&\left( e^{ik}+e^{-ik}+V_{\beta }+e^{-i\phi _{\beta }}\right) f_{0}.
\end{eqnarray}%
For the left incidence ($D=0$), we obtain $r_{\mathrm{L}}=B/A$ and $t_{%
\mathrm{L}}=C/A$ as
\begin{eqnarray}
r_{\mathrm{L}} &=&\frac{\left( e^{-ik}+e^{i\phi _{\alpha }}\right) \frac{%
\left( e^{ik}+e^{-ik}+V_{\alpha }+e^{-i\phi _{\alpha }}\right) \left(
e^{-ik}+V_{\beta }\right) }{\left( e^{ik}+e^{-ik}+V_{\beta }+e^{-i\phi
_{\beta }}\right) }-\left[ \frac{\left( e^{ik}+e^{-ik}+e^{i\phi _{\alpha
}}+e^{i\phi _{\beta }}\right) \left( e^{-ik}+V_{\beta }\right) }{\left(
e^{ik}+e^{-ik}+V_{\beta }+e^{-i\phi _{\beta }}\right) }-\left(
e^{ik}+e^{i\phi _{\beta }}\right) \right] \left( e^{ik}+V_{\alpha }\right) }{%
\left[ \frac{\left( e^{ik}+e^{-ik}+e^{i\phi _{\alpha }}+e^{i\phi _{\beta
}}\right) \left( e^{-ik}+V_{\beta }\right) }{\left( e^{ik}+e^{-ik}+V_{\beta
}+e^{-i\phi _{\beta }}\right) }-\left( e^{ik}+e^{i\phi _{\beta }}\right) %
\right] \left( e^{-ik}+V_{\alpha }\right) -\left( e^{ik}+e^{i\phi _{\alpha
}}\right) \frac{\left( e^{ik}+e^{-ik}+V_{\alpha }+e^{-i\phi _{\alpha
}}\right) \left( e^{-ik}+V_{\beta }\right) }{\left( e^{ik}+e^{-ik}+V_{\beta
}+e^{-i\phi _{\beta }}\right) }},  \label{PTRL} \\
t_{\mathrm{L}} &=&\frac{\left[ \left( e^{-ik}+e^{i\phi _{\alpha }}\right)
-\left( e^{ik}+e^{i\phi _{\alpha }}\right) \frac{\left( e^{ik}+V_{\alpha
}\right) }{\left( e^{-ik}+V_{\alpha }\right) }\right] \left(
e^{-ik}+V_{\alpha }\right) }{\left[ \frac{\left( e^{ik}+e^{-ik}+e^{i\phi
_{\alpha }}+e^{i\phi _{\beta }}\right) \left( e^{-ik}+V_{\beta }\right) }{%
\left( e^{ik}+e^{-ik}+V_{\beta }+e^{-i\phi _{\beta }}\right) }-\left(
e^{ik}+e^{i\phi _{\beta }}\right) \right] \left( e^{-ik}+V_{\alpha }\right)
-\left( e^{ik}+e^{i\phi _{\alpha }}\right) \frac{\left(
e^{ik}+e^{-ik}+V_{\alpha }+e^{-i\phi _{\alpha }}\right) \left(
e^{-ik}+V_{\beta }\right) }{\left( e^{ik}+e^{-ik}+V_{\beta }+e^{-i\phi
_{\beta }}\right) }}.  \label{PTTL}
\end{eqnarray}%
For the right incidence ($A=0$), we obtain $r_{\mathrm{R}}=C/D$ and $t_{%
\mathrm{R}}=B/D$ as%
\begin{eqnarray}
r_{\mathrm{R}} &=&\frac{\frac{\left( e^{ik}+e^{-ik}+V_{\beta }+e^{-i\phi
_{\beta }}\right) \left( e^{-ik}+V_{\alpha }\right) }{\left(
e^{ik}+e^{-ik}+V_{\alpha }+e^{-i\phi _{\alpha }}\right) }\left(
e^{-ik}+e^{i\phi _{\beta }}\right) -\left[ \frac{\left(
e^{ik}+e^{-ik}+e^{i\phi _{\alpha }}+e^{i\phi _{\beta }}\right) \left(
e^{-ik}+V_{\alpha }\right) }{\left( e^{ik}+e^{-ik}+V_{\alpha }+e^{-i\phi
_{\alpha }}\right) }-\left( e^{ik}+e^{i\phi _{\alpha }}\right) \right]
\left( e^{ik}+V_{\beta }\right) }{\left[ \frac{\left(
e^{ik}+e^{-ik}+e^{i\phi _{\alpha }}+e^{i\phi _{\beta }}\right) \left(
e^{-ik}+V_{\alpha }\right) }{\left( e^{ik}+e^{-ik}+V_{\alpha }+e^{-i\phi
_{\alpha }}\right) }-\left( e^{ik}+e^{i\phi _{\alpha }}\right) \right]
\left( e^{-ik}+V_{\beta }\right) -\frac{\left( e^{ik}+e^{-ik}+V_{\beta
}+e^{-i\phi _{\beta }}\right) \left( e^{-ik}+V_{\alpha }\right) }{\left(
e^{ik}+e^{-ik}+V_{\alpha }+e^{-i\phi _{\alpha }}\right) }\left(
e^{ik}+e^{i\phi _{\beta }}\right) },  \label{PTRR} \\
t_{\mathrm{R}} &=&\frac{\left[ \left( e^{-ik}+e^{i\phi _{\beta }}\right)
-\left( e^{ik}+e^{i\phi _{\beta }}\right) \frac{\left( e^{ik}+V_{\beta
}\right) }{\left( e^{-ik}+V_{\beta }\right) }\right] \left( e^{-ik}+V_{\beta
}\right) }{\left[ \frac{\left( e^{ik}+e^{-ik}+e^{i\phi _{\alpha }}+e^{i\phi
_{\beta }}\right) \left( e^{-ik}+V_{\alpha }\right) }{\left(
e^{ik}+e^{-ik}+V_{\alpha }+e^{-i\phi _{\alpha }}\right) }-\left(
e^{ik}+e^{i\phi _{\alpha }}\right) \right] \left( e^{-ik}+V_{\beta }\right) -%
\frac{\left( e^{ik}+e^{-ik}+V_{\beta }+e^{-i\phi _{\beta }}\right) \left(
e^{-ik}+V_{\alpha }\right) }{\left( e^{ik}+e^{-ik}+V_{\alpha }+e^{-i\phi
_{\alpha }}\right) }\left( e^{ik}+e^{i\phi _{\beta }}\right) }.  \label{PTTR}
\end{eqnarray}%
After substituting $V_{\alpha }=-e^{-ik}$ and $V_{\beta }=-e^{ik}$ into the
expressions, we obtain the scattering coefficients $r_{\mathrm{L}}$, $t_{%
\mathrm{L}}$,\ $r_{\mathrm{R}}$, and $t_{\mathrm{R}}$ after simplification.

The left reflection coefficient $r_{\mathrm{L}}$ diverges at $\phi _{\alpha
}=\pi \pm k$ when $\phi _{\beta }\neq \pi \pm k$. At $r_{\mathrm{L}}$
divergence of $\phi _{\alpha }=\pi +k$, we obtain $t_{\mathrm{R}}\rightarrow
\infty $ and%
\begin{eqnarray}
t_{\mathrm{L}} &=&-\frac{\left( e^{ik}-e^{-ik}\right) ^{2}\left(
e^{-ik}+e^{-i\phi _{\beta }}\right) }{2\left( \cos k+\cos \phi _{\beta
}-4i\sin ^{3}k\right) }, \\
r_{\mathrm{R}} &=&\frac{e^{-2ik}\left( \cos \phi _{\beta }+\cos k\right) }{%
\cos \phi _{\beta }+\cos k-4i\sin ^{3}k}.
\end{eqnarray}%
At $r_{\mathrm{L}}$ divergence of $\phi _{\alpha }=\pi -k$, we obtain $t_{%
\mathrm{L}}\rightarrow \infty $ and%
\begin{eqnarray}
t_{\mathrm{R}} &=&-\frac{\left( e^{ik}-e^{-ik}\right) ^{2}\left(
e^{-ik}+e^{i\phi _{\beta }}\right) }{2\left( \cos k+\cos \phi _{\beta
}-4i\sin ^{3}k\right) }, \\
r_{\mathrm{R}} &=&\frac{e^{-2ik}\left( \cos \phi _{\beta }+\cos k\right) }{%
\cos \phi _{\beta }+\cos k-4i\sin ^{3}k}.
\end{eqnarray}%
Notice that $r_{\mathrm{R}}$ are identical for $r_{\mathrm{L}}$ divergence
at $\phi _{\alpha }=\pi \pm k$.

For $\phi _{\alpha }=\pi \pm k$ and $\phi _{\beta }=\pi \pm k$. The
scattering coefficients $\left\{ r_{\mathrm{L}},t_{\mathrm{L}},r_{\mathrm{R}%
},t_{\mathrm{R}}\right\} $ are $\left\{ -1,1,0,-1\right\} $ at (a) $\phi
_{\alpha }=\pi +k,\phi _{\beta }=\pi -k$; $\left\{ -1,\infty ,0,0\right\} $
at (b) $\phi _{\alpha }=\pi -k,\phi _{\beta }=\pi -k$; $\left\{
-1,0,0,\infty \right\} $ at (c) $\phi _{\alpha }=\pi +k,\phi _{\beta }=\pi
+k $; and $\left\{ -1,-1,0,1\right\} $ at (d) $\phi _{\alpha }=\pi -k,\phi
_{\beta }=\pi +k$. The dynamics are illustrated in Fig.~\ref%
{figPTdynamics}.

We can also directly substitute $V_{\alpha }=-e^{-ik}$ and $V_{\beta
}=-e^{ik}$ into the equations of motion at steady-state [Eqs.$~$(\ref{Two1}-%
\ref{Two5})], the calculation of the scattering coefficients is more concise
(however, the scattering coefficients at $r_{\mathrm{L}}$ divergence can not
be obtained in this way). We can obtain
\begin{eqnarray}
f_{\alpha } &=&A+B-f_{0}, \\
e^{ik}f_{\alpha } &=&\left( Ae^{-ik}+Be^{ik}\right) +e^{-i\phi _{\alpha
}}f_{0}, \\
\left( e^{ik}+e^{-ik}\right) f_{0} &=&\left( Ae^{-ik}+Be^{ik}\right) +\left(
Ce^{ik}+De^{-ik}\right) +e^{i\phi _{\alpha }}f_{\alpha }+e^{i\phi _{\beta
}}f_{\beta }, \\
e^{-ik}f_{\beta } &=&\left( Ce^{ik}+De^{-ik}\right) +e^{-i\phi _{\beta
}}f_{0}, \\
f_{\beta } &=&C+D-f_{0},
\end{eqnarray}%
then, the scattering coefficients when $r_{\mathrm{L}}$ does not diverge ($%
\phi _{\alpha }\neq \pi \pm k$) can be calculated as follows. For the left
incidence ($D=0$), we obtain%
\begin{eqnarray}
r_{\mathrm{L}} &=&\frac{B}{A}=\frac{\left( e^{ik}+e^{i\phi _{\beta }}\right)
\left( e^{ik}+e^{-i\phi _{\beta }}\right) -\left( e^{-ik}+e^{i\phi _{\alpha
}}\right) \left( e^{-ik}+e^{-i\phi _{\alpha }}\right) }{\left( e^{i\phi
_{\alpha }}+e^{ik}\right) \left( e^{-i\phi _{\alpha }}+e^{ik}\right) }, \\
t_{\mathrm{L}} &=&\frac{C}{A}=-\frac{e^{-i\phi _{\beta }}+e^{-ik}}{e^{-i\phi
_{\alpha }}+e^{ik}}.
\end{eqnarray}%
For the right incidence ($A=0$), we obtain
\begin{eqnarray}
r_{\mathrm{R}} &=&\frac{C}{D}=0, \\
t_{\mathrm{R}} &=&\frac{B}{D}=-\frac{e^{i\phi _{\beta }}+e^{-ik}}{e^{i\phi
_{\alpha }}+e^{ik}}.
\end{eqnarray}%
The coefficients acquired are in accord with Eqs.~(\ref{PTRL}-\ref{PTTR}).

We perform the time evolution of a Gaussian wave packet to demonstrate the
scattering dynamics, the uniformly coupled resonator chain is cut at the
resonators $-100$ and $100$. The Gaussian wave packet used in the
simulations is
\begin{equation}
\left\vert \Psi \left( 0,j\right) \right\rangle =(\sqrt{\pi }/\sigma
)^{-1/2}\sum_{j}e^{-(\sigma ^{2}/2)\left( j-N_{\mathrm{c}}\right)
^{2}}e^{ik_{\mathrm{c}}j}\left\vert j\right\rangle ,
\end{equation}%
centered at $N_{\mathrm{c}}$ with wave vector $k_{\mathrm{c}}$, and $\sigma $
characterizes its width. $j$ is the resonator index and $\left\vert
j\right\rangle $ is the basis of the resonator chain.

\begin{figure}[b]
\centering%
\includegraphics[ bb=0 0 540 245, width=17.8 cm,
clip]{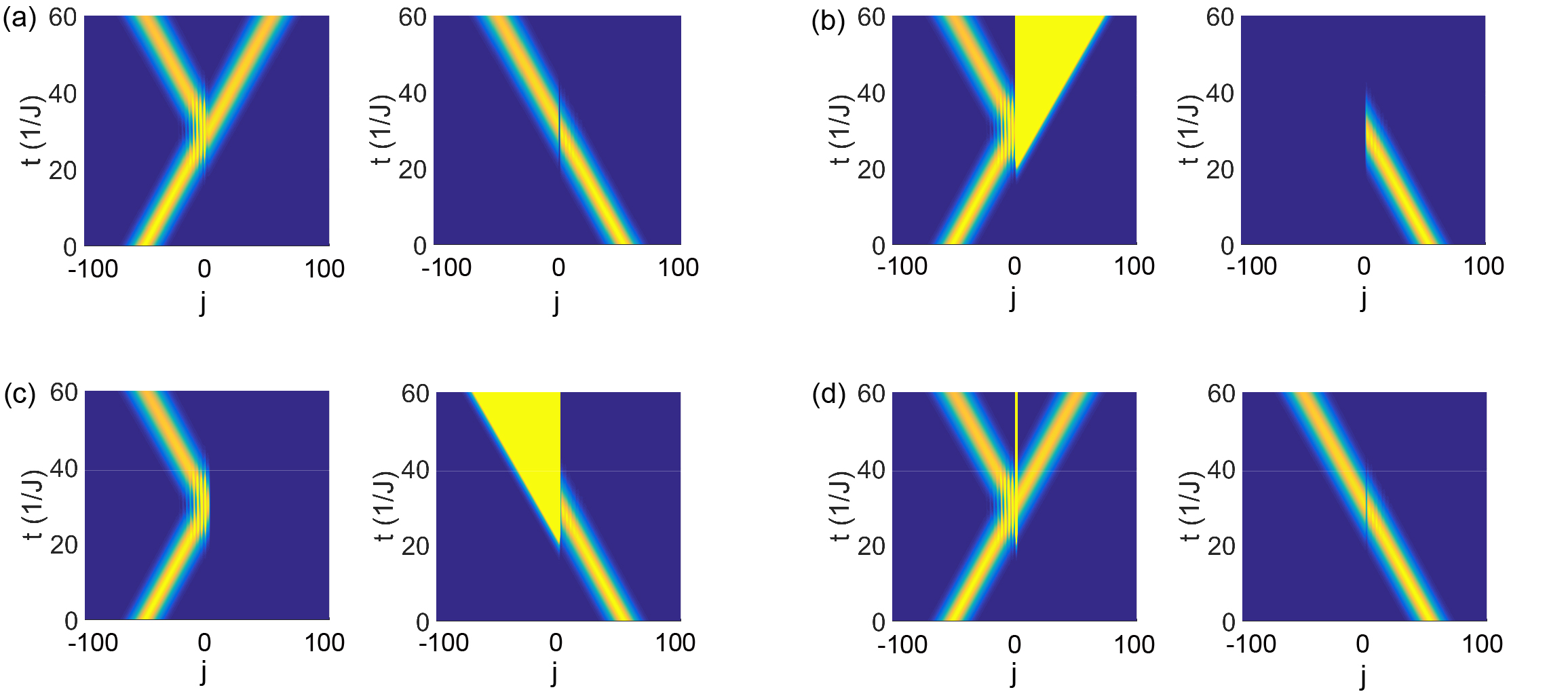}
\caption{Snapshots of the Gaussian wave packet dynamics for the left and
right incidences at (a) $\protect\phi _{\protect\alpha }=-\protect\phi _{\protect\beta }=-2\protect\pi /3$, (b) $\protect\phi _{\protect\alpha }=\protect\phi _{\protect\beta }=2\protect\pi /3$, (c) $\protect\phi _{\protect\alpha }=\protect\phi _{\protect\beta }=-2\protect\pi /3$, and (d) $\protect\phi _{\protect\alpha }=-\protect\phi _{\protect\beta }=2\protect\pi /3$.
They correspond to the triangle, circle, square, and diamond marked in Fig.~3(a) in the Letter. The
parameters are $V_{\protect\alpha }=-e^{-i\protect\pi /3}$, $V_{\protect\beta }=-e^{i\protect\pi /3}$, and $V_{0}=0$. The Gaussian wave packet with $\protect\sigma =0.1$ has wave vector $k_{\mathrm{c}}=\protect\pi /3$ in the
simulations. $\left\vert \left\vert \Psi \left( t,j\right)
\right\rangle \right\vert ^{2}$ is depicted.} \label{figPTdynamics}
\end{figure}

The dynamics at special cases of $\left\vert \phi _{\alpha }\right\vert
=2\pi /3$ and $\left\vert \phi _{\beta }\right\vert =2\pi /3$ are depicted
in Fig.~\ref{figPTdynamics}. Synthetic magnetic flux at $%
\phi _{\alpha }=-2\pi /3$ ($2\pi /3$) produces a left-going (right-going)
wave emission [Fig.~\ref{figUL}]; synthetic magnetic flux at
$\phi _{\beta }=-2\pi /3$ ($2\pi /3$) realizes a right-going (left-going)
wave absorption [Fig.~\ref{figUPA}]. In Figs.~\ref{figPTdynamics}(a) and~\ref{figPTdynamics}(d), the whole
scattering system is $\mathcal{PT}$-symmetric. The unidirectional spectral
singularities for wave emission and wave absorption coincide. $\mathcal{PT}$
symmetry ensures that the persistently emitted waves from resonator $\alpha $
are directly absorbed at resonator $\beta $ and form a unity transmittivity.
The transmittivity is symmetric $|t_{\mathrm{L}}|^2=|t_{\mathrm{R}}|^2=1$,
and the spectral singularities vanish. The snapshots presented in
Figs.~\ref{figPTdynamics}(a) and~\ref{figPTdynamics}(d) are
similar: the transmittivity and reflectivity for the left incidence are both
unity; the transmittivity is unity and reflectivity is zero for the right
incidence. However, the modal amplitude diverges at resonator $0$ when $\phi
_{\alpha }=2\pi /3$ and $\phi _{\beta }=-2\pi /3$, which is imprinted from
the bright line at resonator $0$ after scattering in Fig.~%
\ref{figPTdynamics}(d). In addition, the difference of the synthetic
magnetic fluxes results in a relative phase difference $\pi $ for the
transmitted waves after scattering between the left and right incidences;
however, the reflection coefficients have the same phase. One-way
propagation is again realized through assembling one more side-coupled
resonator. Connecting a unidirectional perfect absorption structure of left
perfect absorption and right resonant transmission [left-UPA in
Fig.~\ref{figUPA}(b) or~\ref{figUPA}(d)] on the right side
of the $\mathcal{PT}$-symmetric two side-coupled resonators at the situation
shown in Fig.~\ref{figPTdynamics}(a) or~\ref{figPTdynamics}%
(d), the one-way propagation is realized as mentioned in section B.

Figure~\ref{figPTdynamics}(b) illustrates a persistent
right-going wave emission (unidirectional lasing) for a left incidence and
perfect absorption for a right incidence. Additional unidirectional perfect
absorption structure of right perfect absorption and left resonant
transmission [right-UPA in Fig.~\ref{figUPA}(a) or~\ref%
{figUPA}(c)] on the left side of the two side-coupled resonators absorbs the
left-going waves; therefore, the unity left reflection is perfectly
absorbed, and the reflectionless left incident unidirectional lasing and
right incident perfect absorption is created as shown in Figs.~4(a)~and~4(b)
in the Letter. The persistent wave emissions are characterized by a Gaussian
error function~\cite{WP}.

Figure~\ref{figPTdynamics}(c) illustrates a persistent
left-going wave emission for a right incidence and a full reflection for a
left incidence; therefore, additional unidirectional perfect absorption
structure of left perfect absorption and right resonant transmission
[left-UPA in Fig.~\ref{figUPA}(b) or~\ref{figUPA}(d)] on the
left side of the two side-coupled resonators absorbs the left incidence, and
the reflectionless right incident unidirectional lasing and left incident
perfect absorption is created as shown in Figs.~4(d) and~4(e) in the Letter.

The scattering coefficients $r_{\mathrm{L,R}}$ $t_{\mathrm{L,R}}$ virtually
share identical denominator, but their numerators are distinct. The system
is at spectral singularities when the denominator goes to zero, provided
that the numerator does not vanish. When the denominator and the numerator
vanish simultaneously, the coefficients are obtained by calculating the
limitation of expressions as wave vector $k$ approaches the divergent wave
vector. The transmission and reflection coefficients do not diverge and the
spectral singularities vanish.

\subsection{Left bidirectional lasing and right perfect absorption}

The influences of the side-coupled resonators $\alpha $, $\beta $ vary as
the synthetic magnetic fluxes at different matches. To demonstrate the
phenomenon of bidirectional lasing from one side and perfect absorption from
the other side, we first show the unidirectional lasing performed in a
Hermitian conjugation system of the unidirectional perfect absorption
illustrated in Figs.~\ref{figUPA}(a) and~\ref{figUPA}(b).
The dynamics for the left and right incidences are depicted in Figs.~\ref{figUL}(a) and~\ref{figUL}(b), respectively. The schematics of
the Hermitian conjugation systems of unidirectional perfect absorption are
illustrated in Fig.~\ref{figUL}(c). Notice that the
unidirectional lasing is toward opposite directions for the opposite
incidences.

\begin{figure}[thb]
\centering\includegraphics[ bb=0 0 390 125, width=17.5 cm,
clip]{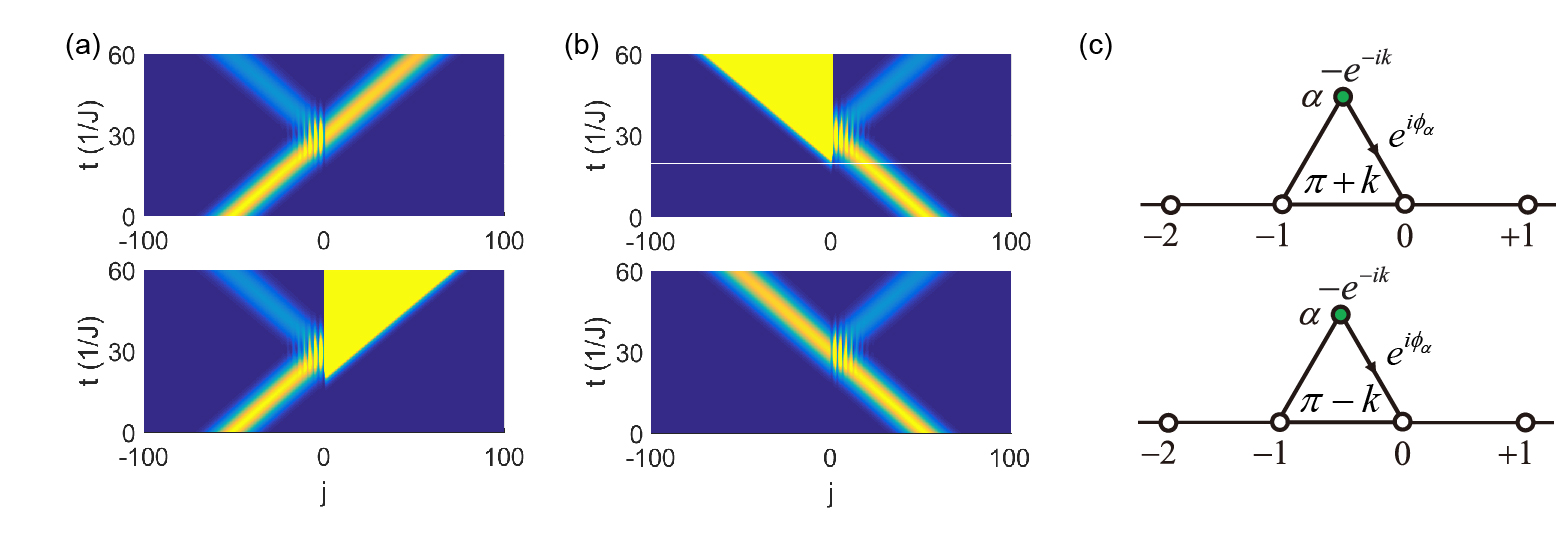}
\caption{(a, b) Snapshots of the Gaussian wave packet dynamics of a unidirectional
lasing at $\protect\phi _{\protect\alpha }=-2\protect\pi /3$ and $\protect\phi _{\protect\alpha }=2\protect\pi /3$ for the upper and lower panels,
respectively. (c) Schematic of the system with parameters $V_{\protect\alpha }=-e^{-i\protect\pi /3}$, $k=\protect\pi/3$. The Gaussian wave packet with $\protect\sigma =0.1$ has wave vector $k_{\mathrm{c}}=\protect\pi /3$ in the
simulations. $\left\vert \left\vert \Psi \left( t,j\right)
\right\rangle \right\vert ^{2}$ is depicted.} \label{figUL}
\end{figure}

\begin{figure}[thb]
\centering\includegraphics[ bb=0 0 390 110, width=17.5 cm,
clip]{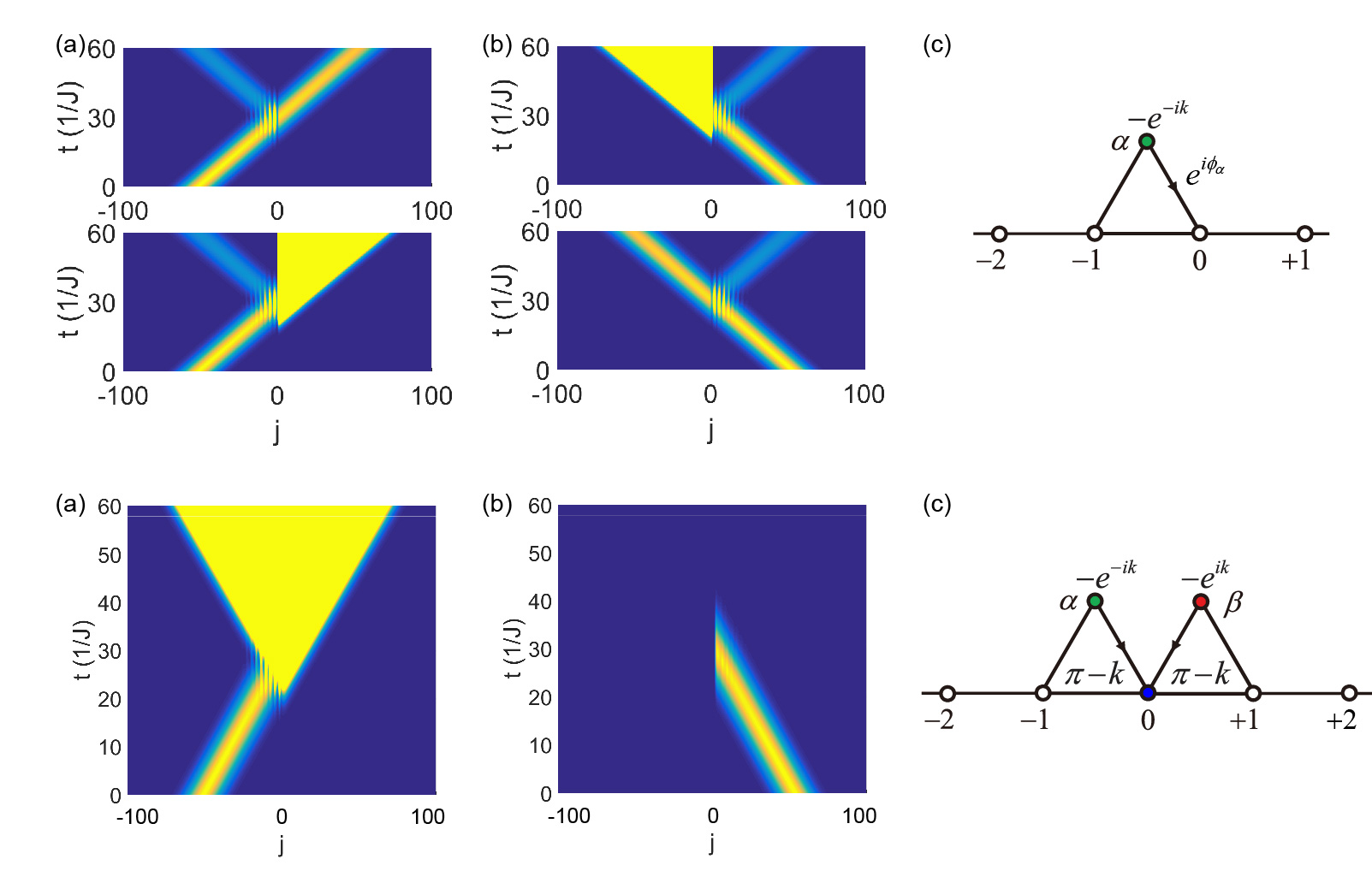}
\caption{Snapshots of the Gaussian wave packet dynamics for the (a) left and
(b) right incidences at $\protect\phi _{\protect\alpha }=\protect\phi _{\protect\beta }=2\protect\pi /3$. (c) Schematic of the system with
parameters $V_{\protect\alpha }=-e^{-i\protect\pi /3}$, $V_{\protect\beta }=-e^{i\protect\pi /3}$, and $V_{0}=-2i\sin \left( \protect\pi /3\right) $, $k=\protect\pi/3$. The Gaussian wave packet with $\protect\sigma =0.1$ has
wave vector $k_{\mathrm{c}}=\protect\pi /3$ in the simulations. $\left\vert \left\vert \Psi \left( t,j\right) \right\rangle
\right\vert ^{2}$ is depicted.} \label{figLaserAbsorber}
\end{figure}

In Fig.~\ref{figLaserAbsorber}, the chosen synthetic
magnetic fluxes are opposite as the situation of a single-direction lasing
for the CCW mode [Figs.~2(c) and~2(d) in the Letter], and other system
parameters are unchanged. That is $\phi _{\alpha }=\phi _{\beta }=\pi -k$, $%
V_{\alpha }=-e^{-ik}$, $V_{0}=-2i\sin k$, and $V_{\beta }=-e^{ik}$; these
characterize the CW mode in the system when the CCW mode is at
single-direction lasing. The phenomenon of bidirectional lasing for
incidence from one direction and perfect absorption for incidence from the
other direction is simulated for a Gaussian wave packet excitation. For a
left incidence, a right-going unidirectional lasing is generated at the gain
resonator $\alpha $ [lower panel of Fig.~\ref{figUL}(a)];
which is scattered at resonator $0$ with half reflected and half
transmitted. The half reflected unidirectional lasing passes through
resonator $\alpha $ from the right side with unity transmittivity [lower
panel of Fig.~\ref{figUL}(b)] and forms the left-going wave
emission; the half transmitted unidirectional lasing resonantly passes
through resonator $\beta $ from its left side and forms the right-going wave
emission; therefore, a symmetric bidirectional lasing is created for the
left incidence. For a right incidence, the Gaussian wave packet is perfectly
absorbed without reflection. At $\phi _{\alpha }=\phi _{\beta }=\pi -k$, $%
V_{\alpha }=-e^{-ik}$, and $V_{\beta }=-e^{ik}$, the bidirectional lasing
for the left incidence is tuned by $V_{0}$ and becomes unidirectional at $%
V_{0}=0$ as depicted in Fig.~\ref{figPTdynamics}(b).

\subsection{Wave propagation dynamics in the presence of backscattering}

The surface roughness and defects in the resonator induce backscattering
between the CCW and CW modes~\cite{OE,Wiersig2016,Morichetti}, which results
in a mode coupling and an intensity interchange between two modes. For the
unidirectional perfect absorption structure in Fig.~\ref%
{figUPA}(a), the equations of motion in the presence of backscattering for
the resonators $j<-1$ and $j>0$ in the chain are revised to
\begin{equation}
i\frac{\mathrm{d}\psi _{\mathrm{CCW},j}}{\mathrm{d}t}=\omega _{\mathrm{c}%
}\psi _{\mathrm{CCW},j}-J\psi _{\mathrm{CCW},j-1}-J\psi _{\mathrm{CCW}%
,j+1}-\kappa _{j}\psi _{\mathrm{CW},j},  \label{CCW1}
\end{equation}%
and%
\begin{equation}
i\frac{\mathrm{d}\psi _{\mathrm{CW},j}}{\mathrm{d}t}=\omega _{\mathrm{c}%
}\psi _{\mathrm{CW},j}-J\psi _{\mathrm{CW},j-1}-J\psi _{\mathrm{CW}%
,j+1}-\kappa _{j}\psi _{\mathrm{CCW},j},
\end{equation}%
where $\psi _{\mathrm{CCW},j}$ and $\psi _{\mathrm{CW},j}$ are the mode
amplitudes of the CCW and CW modes in the resonator $j$, respectively. The
coupling between the CCW and CW modes of the resonator $j$ is $\kappa _{j}$~%
\cite{TJK}. The equations of motion for other resonators $j=-1$, $0$, and $%
\alpha $ are%
\begin{eqnarray}
&&i\frac{\mathrm{d}\psi _{\mathrm{CCW,}-1}}{\mathrm{d}t}=\omega _{\mathrm{c}%
}\psi _{\mathrm{CCW,}-1}-J\psi _{\mathrm{CCW,}-2}-J\psi _{\mathrm{CCW,}%
0}-g\psi _{\mathrm{CCW,}\alpha }-\kappa _{-1}\psi _{\mathrm{CW,}-1}, \\
&&i\frac{\mathrm{d}\psi _{\mathrm{CCW,}\alpha }}{\mathrm{d}t}=\left( \omega
_{\mathrm{c}}+V_{\alpha }\right) \psi _{\mathrm{CCW,}\alpha }-gJ\psi _{%
\mathrm{CCW,}-1}-ge^{-i\phi _{\alpha }}\psi _{\mathrm{CCW,}0}-\kappa
_{\alpha }\psi _{\mathrm{CW,}\alpha }, \\
&&i\frac{\mathrm{d}\psi _{\mathrm{CCW,}0}}{\mathrm{d}t}=\omega _{\mathrm{c}%
}\psi _{\mathrm{CCW,}0}-J\psi _{\mathrm{CCW,}-1}-J\psi _{\mathrm{CCW,}%
1}-ge^{i\phi _{\alpha }}\psi _{\mathrm{CCW,}\alpha }-\kappa _{0}\psi _{%
\mathrm{CW,}0},
\end{eqnarray}%
and
\begin{eqnarray}
&&i\frac{\mathrm{d}\psi _{\mathrm{CW,}-1}}{\mathrm{d}t}=\omega _{\mathrm{c}%
}\psi _{\mathrm{CW,}-1}-J\psi _{\mathrm{CW,}-2}-J\psi _{\mathrm{CW,}0}-g\psi
_{\mathrm{CW,}\alpha }-\kappa _{-1}\psi _{\mathrm{CCW,}-1}, \\
&&i\frac{\mathrm{d}\psi _{\mathrm{CW,}\alpha }}{\mathrm{d}t}=\left( \omega _{%
\mathrm{c}}+V_{\alpha }\right) \psi _{\mathrm{CW,}\alpha }-gJ\psi _{\mathrm{%
CW,}-1}-ge^{-i\phi _{\alpha }}\psi _{\mathrm{CW,}0}-\kappa _{\alpha }\psi _{%
\mathrm{CCW,}\alpha }, \\
&&i\frac{\mathrm{d}\psi _{\mathrm{CW,}0}}{\mathrm{d}t}=\omega _{\mathrm{c}%
}\psi _{\mathrm{CW,}0}-J\psi _{\mathrm{CW,}-1}-J\psi _{\mathrm{CW,}%
1}-ge^{i\phi _{\alpha }}\psi _{\mathrm{CW,}\alpha }-\kappa _{0}\psi _{%
\mathrm{CCW,}0}.
\end{eqnarray}

In the simulations shown below, the side-coupling is set $g=J=1$ as that in
the Letter. The time evolution intensity $\left\vert \left\vert \Psi \left(
t,j\right) \right\rangle \right\vert ^{2}$ of an initial Gaussian excitation
$\left\vert \Psi \left( 0,j\right) \right\rangle =(\sqrt{\pi }/\sigma
)^{-1/2}\sum_{j}e^{-(\sigma ^{2}/2)\left( j-N_{\mathrm{c}}\right)
^{2}}e^{ik_{\mathrm{c}}j}\left\vert j\right\rangle $ is depicted, $\sigma $
characterizes its width. The initial Gaussian wave packet CCW mode
excitation is centered at $N_{\mathrm{c}}$ with wave vector $k_{\mathrm{c}}$%
, the velocity of the Gaussian wave packet is $2J\sin \left( k_{\mathrm{c}%
}\right) $. The mode coupling between the CCW and CW modes results in an
intensity interchange between these two modes in the wave propagation
process, which can be characterized by an intensity breathing~\cite{LJin}.

\begin{figure}[thb]
\centering\includegraphics[ bb=0 0 600 490, width=17.5 cm,
clip]{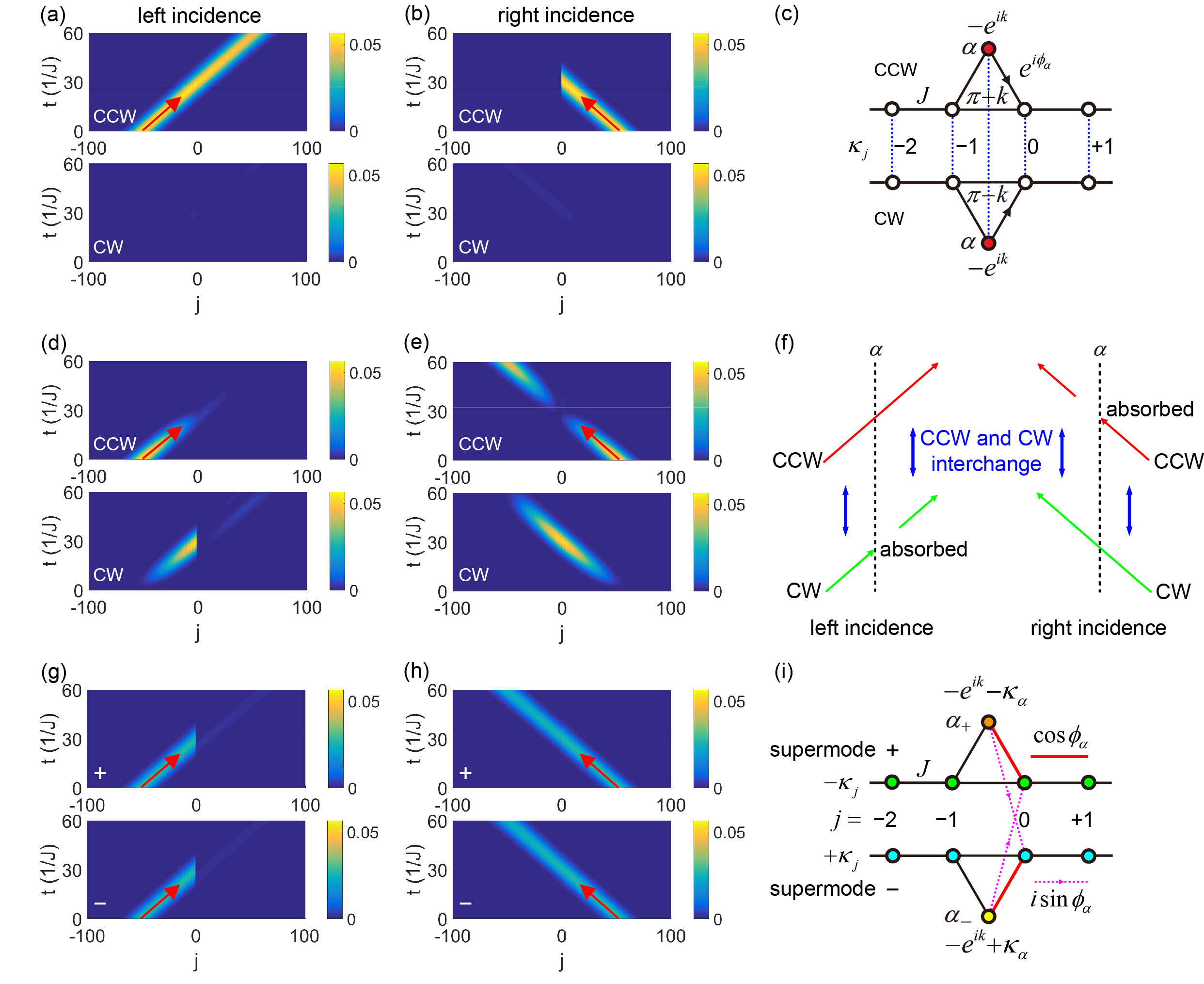}
\caption{(a, b, d, e) Snapshots of the intensity of a unidirectional perfect
absorption. The upper (lower) panels are for the CCW (CW) mode. The initial
excitation is the CCW mode Gaussian wave packet with $\protect\sigma =0.1$ and wave vector $k_{\mathrm{c}}=\protect\pi /3$. The red arrows indicate the
incidences. (c) Schematic of the system in the presence of backscattering,
the coupling between the CCW and CW modes is indicated by the dotted blue
lines. The system parameters are marked, $V_{\protect\alpha }=-e^{-i k}$, $k=\protect\pi/3$. The phase direction of the coupling is indicated by the
arrows in the triangular structures, the values of opposite synthetic magnetic fluxes experienced by the
CCW and CW modes are marked. (f) Cartoon of the wave propagation in the
presence of backscattering. (g, h) Snapshots of the intensity of a supermode unidirectional perfect
absorption, the parameters are identical with that in (d, e).
The upper (lower) panels are for the symmetric (antisymmetric) supermode $+$ ($-$). (i) Schematic of the system in the supermode basis, $\phi_{\alpha}=\pi+k$. The green (cyan) circles indicate the symmetric (antisymmetric) supermode basis with frequency $\omega_{\rm{c}}-\kappa_{j}$  ($\omega_{\rm{c}}+\kappa_{j}$).
In the simulations, each $\protect\kappa_j$
is randomly chosen within $[0,0.01J]$ in (a, b), within $[0,0.1J]$ in (d, e, g, h), and $J=1$; the wave propagation dynamics are averaged over $10^3$ sample systems. Notice that the initial intensities in both the upper and lower panels of (g, h) are half that in the upper panels of (a, b).
} \label{figUPABS}
\end{figure}

In Figs.~\ref{figUPABS}(a) and~\ref{figUPABS}(b), the
dynamics of a unidirectional perfect absorption are depicted in the presence
of backscattering. Schematic of an equivalent configuration is shown in
Fig.~\ref{figUPABS}(c). At weak mode coupling ($0\leqslant
\kappa _{j}\leqslant 0.01J$), the unidirectional perfect absorption
performed for the CCW mode excitation remains good.

In Figs.~\ref{figUPABS}(d) and~\ref{figUPABS}(e), the
coupling between the CCW and CW modes in each resonator is randomly chosen
within the region $0\leqslant $ $\kappa _{j}\leqslant 0.1J$. A deviation
from an ideal unidirectional perfect absorption is noticed. For a left
incident CCW mode excitation [Fig.~\ref{figUPABS}(d)], a CW
mode Gaussian wave packet is formed with an identical velocity of the CCW
mode when it propagates toward resonator $\alpha $; the CCW mode wave packet
resonantly transmits at resonator $\alpha $, but the CW mode wave packet is
perfectly absorbed as predicted by Eq.~(\ref{S}). After the CCW mode wave
packet resonantly passing through resonator $\alpha $, the CW mode is formed
once again through the mode coupling (i.e., mode intensity breathing).
Similarly, for a right incident CCW mode excitation [Fig.~%
\ref{figUPABS}(e)], the incident left-going CCW mode wave packet is
perfectly absorbed at resonator $\alpha $; the backscattering induced CW
mode wave packet resonantly passes through resonator $\alpha $ from the
right side to the left side, which transfers into the CCW mode after
scattering. These dynamical processes are schematically illustrated in
Fig.~\ref{figUPABS}(f), where the red arrows represent the
CCW mode excitations, the blue double-arrows represent the interchange
between the CCW and CW modes, and the green arrows represent the CW mode
formed through mode coupling in the wave propagation process of the CCW mode
excitation.

The mode coupling $\kappa _{j}$ mixes the CCW and CW modes in the resonator $%
j$, results in the frequency shift and mode splitting~\cite{Wiersig2016},
and creates symmetric and antisymmetric supermodes $\left\vert j,\pm
\right\rangle =\left( \left\vert j,\mathrm{CCW}\right\rangle \pm \left\vert
j,\mathrm{CW}\right\rangle \right) /\sqrt{2}$ with resonant frequencies $%
\omega _{\mathrm{c}}\mp \kappa _{j}$~\cite{TJK}, respectively. Fig.~\ref{figUPABS}(i) is a schematic of the unidirectional perfect
absorption structure of Fig.~\ref{figUPABS}(c) after applied
a uniform transformation, the new basis is the supermodes instead of the
\textrm{CCW} and \textrm{CW} modes.\ The supermode splitting $2\kappa $ (if
assuming a constant mode coupling $\kappa =\kappa _{j}$) indicates that\ the
mode coupling induced intensity breathing has a period of $T=\pi /\kappa $~%
\cite{LJin}. We can roughly estimate the condition of obtaining a good
performance unidirectional perfect absorption, which is approximately%
\begin{equation}
N_{\mathrm{c}}/[J\sin \left( k_{\mathrm{c}}\right) ]\ll T.
\end{equation}%
For $k_{\mathrm{c}}=\pi /3$ and $N_{\mathrm{c}}=50$, we have $\kappa \ll \pi
J\sin \left( k_{\mathrm{c}}\right) /N_{\mathrm{c}}\approx 0.05J$. Notice
that the parameters in Figs.~\ref{figUPABS}(a) and~\ref%
{figUPABS}(b) satisfy this condition.

The intensities of the \textrm{CCW} and \textrm{CW} modes $|\left\vert \Psi
\left( t,j\right) \right\rangle _{\mathrm{CCW}}|^{2}$, $|\left\vert \Psi
\left( t,j\right) \right\rangle _{\mathrm{CW}}|^{2}$ are depicted in
Figs.~\ref{figUPABS}(d) and~\ref{figUPABS}(e), notice that
the breathing period approaches the propagating time ($T\approx 63/J$ and $%
N_{\mathrm{c}}/[J\sin \left( k_{\mathrm{c}}\right) ]\approx 58/J$).
Correspondingly, the intensities of the supermodes $|\left( \left\vert \Psi
\left( t,j\right) \right\rangle _{\mathrm{CCW}}\pm \left\vert \Psi \left(
t,j\right) \right\rangle _{\mathrm{CW}}\right) /\sqrt{2}|^{2}$ are depicted
in Figs.~\ref{figUPABS}(g) and~\ref{figUPABS}(h). An
interesting phenomenon occurs in this situation: both the symmetric and
antisymmetric supermodes are left incident perfect absorption and right
incident resonant transmission. This is\textit{\ a consequence of the proper
match of the mode coupling and the wave propagating time} and can be
understand as follows. In Fig.~\ref{figUPABS}(d), it is
noticed that when the \textrm{CCW} mode wave packet reaches resonator $%
\alpha $, it is completely changed into the CW mode due to the proper match
of mode interchange and the wave packet propagating time; then the \textrm{CW%
} mode is perfectly absorbed at resonator $\alpha $. The UPA absorbs the CW
mode without affecting the CCW mode, which helps preventing the unwanted
backscattering induced CW mode accumulation in the left side. This results
in the left perfect absorption for both the symmetric and antisymmetric
supermodes as shown in Fig.~\ref{figUPABS}(g). The right
incident \textrm{CCW} mode is completely changed into the \textrm{CW} mode
when the wave packet reaches resonator $\alpha $, and then escapes from
being absorbed; the \textrm{CW} mode formed resonantly passes through
resonator $\alpha $ and then changes back into the \textrm{CCW} mode as
shown in Fig.~\ref{figUPABS}(e). In this process, the right
incident CCW mode is almost not affected by resonator $\alpha $; therefore,
both the symmetric and antisymmetric mode incidences resonantly transmit at
resonator $\alpha $ as shown in Fig.~\ref{figUPABS}(h).
Here, the unidirectional perfect absorber acts similarly as an isolator for
the supermodes.

\begin{figure}[thb]
\centering\includegraphics[ bb=0 0 600 170, width=17.5 cm,
clip]{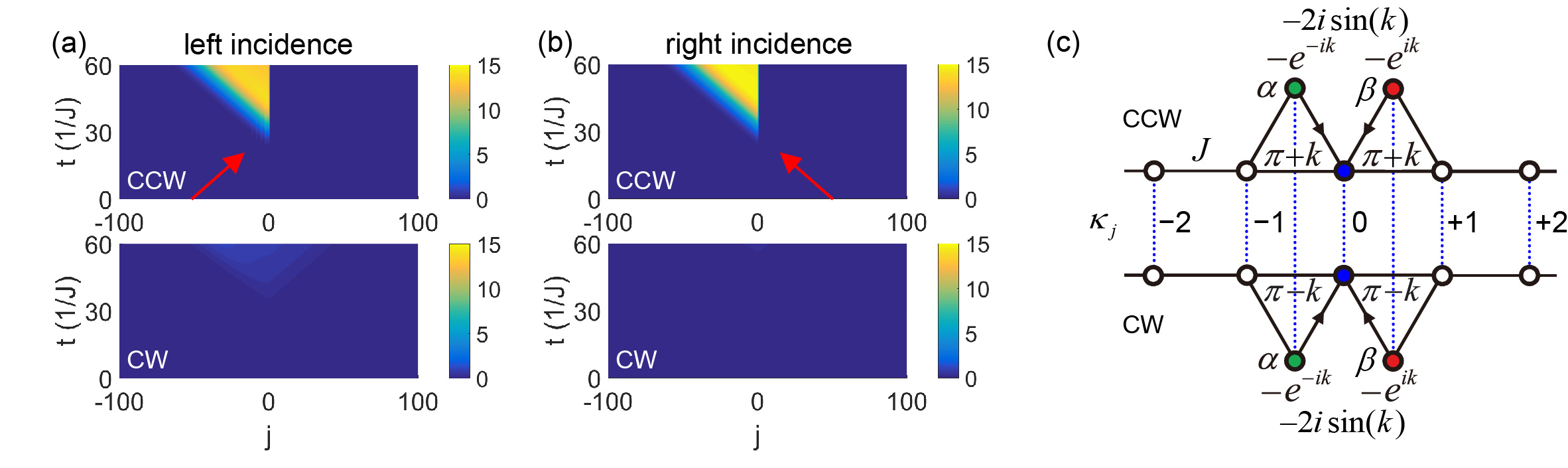}
\caption{(a, b) Snapshots of the intensity of a single-direction lasing. The
upper (lower) panels are for the CCW (CW) mode. The initial excitation is
the CCW mode Gaussian wave packet with $\protect\sigma =0.1$ and wave vector
$k_{\mathrm{c}}=\protect\pi /3$. The red arrows indicate the incidences. (c) Schematic of the
system in the presence of backscattering, the coupling between the CCW and
CW modes is indicated by the dotted blue lines. The system parameters are
marked, $k=\protect\pi/3$. The coupling phase direction is indicated by the
arrows in the triangular structures, which reflect the opposite synthetic magnetic fluxes that
experienced by the CCW and CW modes. In the simulations, each $\protect\kappa_j$ is randomly chosen within $[0,0.01J]$, $J=1$; the wave propagation dynamics are averaged
over $10^3$ sample systems.} \label{DL}
\end{figure}

In the presence of backscattering, the equations of motion for the systems
with more side-coupled resonators are similarly extended with an additional
term characterizing the coupling between the CCW and CW modes in each
equation. In Figs.~\ref{DL}(a) and~\ref{DL}(b), the dynamics
of a single-direction lasing for the CCW mode excitation is performed at
weak mode coupling ($0\leqslant \kappa _{j}\leqslant 0.01J$). Schematic of
an equivalent configuration is shown in Fig.~\ref{DL}(c).
Notably, the single-direction lasing depicted in the simulation remains
good. The CW mode, formed in the wave propagation process due to the mode
coupling, induces a weak bidirectional lasing (Fig.~\ref%
{figLaserAbsorber}) in the lower panel of Fig.~\ref{DL}(a)
for the left incident CCW mode excitation, and the CW mode formed in the
wave propagation process is perfectly absorbed for the right incident CCW
mode excitation as shown in the lower panel of Fig.~\ref{DL}%
(b).

\begin{figure}[tb]
\centering\includegraphics[ bb=20 0 600 160, width=17.5 cm,
clip]{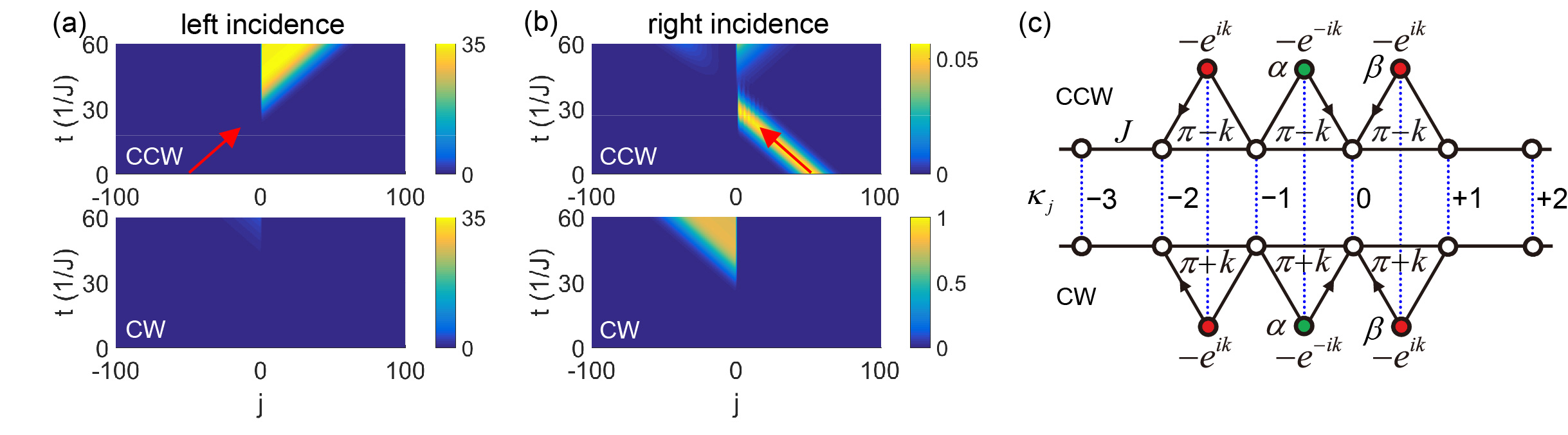}
\caption{Snapshots of the intensity of a reflectionless unidirectional lasing and
perfect absorption for the incidences from the (a) left and (b) right. The upper (lower)
panels are for the CCW (CW) mode. The initial excitation is the CCW mode Gaussian wave packet with $\protect\sigma =0.1$ and wave vector $k_{\mathrm{c}}=\protect\pi /3$. The red arrows indicate the incidences.
(c) Schematic of the system in the presence of backscattering, the coupling between the CCW and CW modes is indicated by the dotted blue lines. The system parameters are marked, where $k=\protect\pi/3$. The coupling phase direction is indicated by the arrows in
the triangular structures, indicating the opposite synthetic magnetic fluxes that
experienced by the CCW and CW modes. In the simulations, each $\protect\kappa_j$ is randomly chosen within $[0,0.01J]$, $J=1$; the wave propagation dynamics are averaged
over $10^3$ sample systems.} \label{RUSS}
\end{figure}

In Figs.~\ref{RUSS}(a) and~\ref{RUSS}(b), the dynamics of a
reflectionless unidirectional lasing and perfect absorption is performed at
weak mode coupling. Schematic of an equivalent configuration is shown in
Fig.~\ref{RUSS}(c). An ideal reflectionless unidirectional
lasing and perfect absorption is depicted in Figs.~4(a-f) of the Letter: a
left incident reflectionless transmission divergence ($r_{\mathrm{L}}=0,t_{%
\mathrm{L}}\rightarrow \infty $) and a right incident perfect absorption ($%
r_{\mathrm{R}}=t_{\mathrm{R}}=0$) for the CCW mode incidence, and a left
incident perfect absorption ($r_{\mathrm{L}}=t_{\mathrm{L}}=0$) and a right
incident reflectionless transmission divergence ($r_{\mathrm{R}}=0,t_{%
\mathrm{R}}\rightarrow \infty $) for the CW mode incidence. Notably, the
reflectionless unidirectional lasing and perfect absorption is severely
affected by the backscattering.
This is because that the mode coupling induced weak CW mode is amplified at
the gain resonator after scattering, which is not properly absorbed at the
dissipative resonators due to the mode coupling; in particular, a weak
left-going CW mode unidirectional lasing is created for a right incident CCW
mode excitation as illustrated in the lower panel of Fig.~%
\ref{RUSS}(b) as predicted in Figs.~4(d-f) of the Letter.

\begin{figure}[thb]
\centering\includegraphics[ bb=10 0 600 170, width=17.5 cm,
clip]{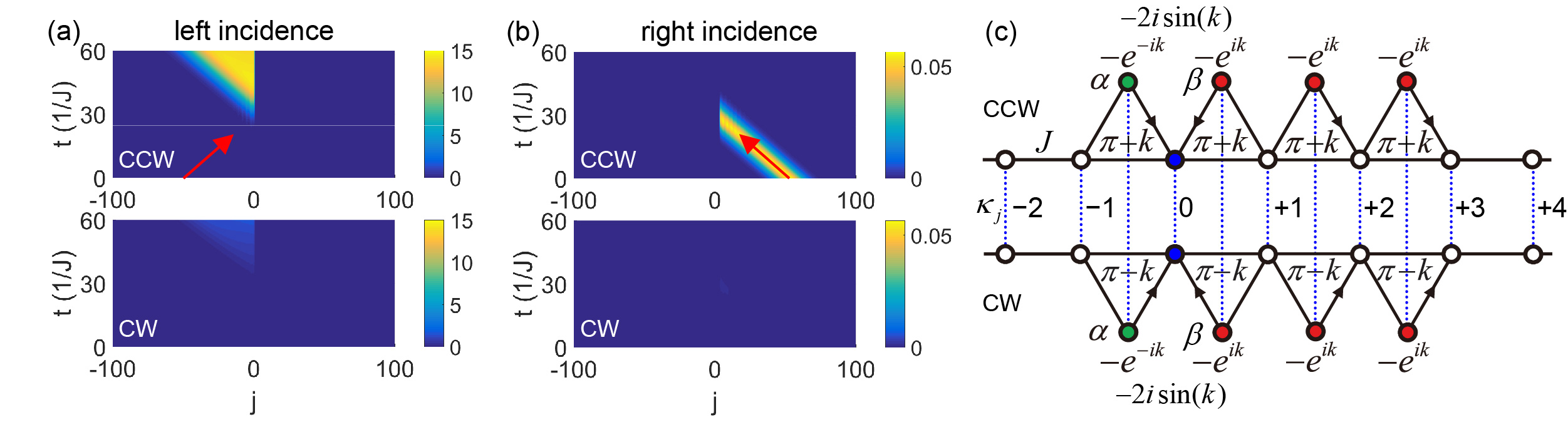}
\caption{Snapshots of the intensity of a transmissionless unidirectional lasing
and perfect absorption for the incidences from the (a) left and (b) right. The upper (lower)
panels are for the CCW (CW) mode. The initial excitation is the CCW mode of
Gaussian wave packet with $\protect\sigma =0.1$ and wave vector $k_{\mathrm{c}}=\protect\pi /3$. The red arrows indicate the incidences. (c) Schematic of the system in the
presence of backscattering, the coupling between the CCW and CW modes is
indicated by the dotted blue lines. The system parameters are marked, $V_{0}=-2i\sin \left( \protect\pi /3\right) $, $k=\protect\pi/3$. The
coupling phase direction is indicated by the arrows in the triangular
structures, which reflect the opposite synthetic magnetic fluxes that experienced by the CCW
and CW modes. In the simulations, each $\protect\kappa_j$ is randomly chosen
within $[0,0.01J]$, $J=1$; the wave propagation dynamics are averaged over $10^3$ sample
systems.} \label{TUSS}
\end{figure}

In Figs.~\ref{TUSS}(a) and~\ref{TUSS}(b), the dynamics of a
transmissionless unidirectional lasing and perfect absorption for the CCW
mode excitation is performed at weak mode coupling ($0\leqslant \kappa
_{j}\leqslant 0.01J$). Schematic of an equivalent configuration is shown in
Fig.~\ref{TUSS}(c). Notably, Fig.~\ref{TUSS}
reveals a good transmissionless unidirectional lasing and perfect absorption
as depicted in Figs.~4(g-i) of the Letter. Both the CCW and CW modes lead to
a simultaneous transmissionless left incident reflection divergence ($t_{%
\mathrm{L}}=0$, $r_{\mathrm{L}}\rightarrow \infty $) and right incident
perfect absorption ($r_{\mathrm{R}}=t_{\mathrm{R}}=0$). The CW mode formed
in the wave propagation process of a left incident CCW mode excitation leads
to a weak left-going unidirectional lasing as depicted in the lower panel of
Fig.~\ref{TUSS}(a). For a right incident CCW mode
excitation, both the left-going CCW mode and the CW mode formed in the wave
propagation process due to the mode coupling are perfectly absorbed as
depicted in the lower panel of Fig.~\ref{TUSS}(b).

\clearpage
\end{widetext}


\begin{thebibliography}{99}
\bibitem{FLUX} The directional phase factor $e^{\pm i\phi _{\alpha ,\beta }}$
in the asymmetric couplings is $2\pi $ periodic of $\phi _{\alpha ,\beta }$.
In our discussions, when the condition $\phi _{\alpha ,\beta }=\pi \pm k$ is
mentioned, we imply that $\phi _{\alpha ,\beta }+2n\pi =\pi \pm k$ ($n\in
\mathbb{Z}
$).
\end{thebibliography}
\end{document}